\documentclass[useAMS, usenatbib]{mn2e}
\usepackage{graphicx}
\usepackage{amsmath}
\usepackage{color}
\usepackage{subfigure}

\title[]{The effect of foreground subtraction on cosmological measurements from Intensity Mapping }
\author[L.Wolz et al.]{L. Wolz$^1$$^,$$^2$, F. B. Abdalla$^1$, C. Blake$^3$, J. R. Shaw$^4$, E. Chapman$^1$, S. Rawlings$^2$\\
$^1$Department of Physics and Astronomy, University College London, London WC1E 6BT, UK\\
$^2$ Sub-Dept. of Astrophysics, Dept. of Physics, University of Oxford,
The Denys Wilkinson Building, Keble Road, Oxford OX1 3RH,
UK\\
$^3$ Centre for Astrophysics \& Supercomputing, Swinburne University of Technology, P.O. Box 218, Hawthorn, VIC 3122, Australia\\
$^4$ Canadian Institute for Theoretical Astrophysics, 60 St. George Street, Toronto, ON M5S 3H8, Canada
}

\newcommand{\recdata}{\rm{cleaned}}
\begin{document}

\date{}
\pagerange{\pageref{firstpage}--\pageref{lastpage}} \pubyear{}
\maketitle
\label{firstpage}
\begin{abstract}
We model a 21 cm intensity mapping survey in the redshift range $0.01<z<1.5$ designed to simulate the skies as seen by future radio telescopes such as the Square Kilometre Array (SKA), including instrumental noise and Galactic foregrounds. 
In our pipeline, we remove the introduced Galactic foregrounds with a fast independent component analysis (\textsc{fastica}) technique. 
We present the power spectrum of the large-scale matter distribution, $C(\ell)$, before and after the application of this foreground removal method and calculate the resulting systematic errors. 
We attempt to reduce systematics in the foreground subtraction by optimally masking the maps to remove high foregrounds in the Galactic plane. 
Our simulations show a certain level of bias remains in the power spectrum at all scales $\ell<400$.
At large-scales $\ell<30$ this bias is particularly significant.
We measure the impact of these systematic effects in two different ways: firstly we fit cosmological parameters to the broadband shape of the power spectrum and secondly we extract the position of the Baryon Acoustic Oscillations (BAO).
In the first analysis, we find that the systematics introduce an significant shift in the best fit cosmological parameters at the 2 to 3 sigma level which depends on the masking and noise levels.
However, cosmic distances can be recovered in an unbiased way after foreground removal at all simulated redshifts by fitting the BAOs in the power spectrum.
We conclude that further advances in foreground removal are needed in order to recover unbiased information from the broadband shape of the power spectrum, however, intensity mapping experiments will be a powerful tool for mapping cosmic distances across a wide redshift range.
 \end{abstract}
\begin{keywords} cosmological parameters, large-scale structure of the Universe, methods: statistical, radio lines: galaxies
\end{keywords}

\section{Introduction}

Cosmological parameters can be estimated via the measurement of the large-scale distribution of the galaxies. This has been successfully performed over the last decade with optical surveys (e.g. \citealt{Percival:2001hw, Tegmark:2003uf, Tegmark:2006az, Blake:2011en, 2011PhRvL.106x1301T}). The accuracy of parameter estimation is significantly increased if the survey covers a large fraction of the sky and can access high redshifts.
However, this is challenging at optical wavelength because integration times are usually long and the high resolution needed to get data from each individual galaxy makes the survey speed slow. 

In most types of survey, the field-of-view (FoV) is a feature of the telescope design and not adjustable during the observations. 
The present and next generation of radio interferometers, such as LOFAR\footnote{\texttt{http://www.lofar.org/}}\citep{vanHaarlem:2013dsa} and the Square Kilometre Array (SKA\footnote{\texttt{http://www.skatelescope.org/}}) have been designed with new technologies such as stations with phased array feeds as well as stations composed of dipoles. This means that surveys with flexible FoV and multibeaming can be easily arranged and performed within much shorter observing times. In this way the FoV of optical surveys, which is typically of the size of $\approx 1\deg^2$, can be greatly exceeded by future radio telescopes.

In the last decade, a technique called intensity mapping (see e.g. \citealt{Chang:2007xk, 2009atnf.prop.2491V, 2008arXiv0807.3614A, Peterson:2009ka, Chang:2010jp, 2012RPPh...75h6901P, 2013ApJ...763L..20M, 2013MNRAS.434L..46S}) has been proposed. The basic idea is that the entire HI flux of a wide patch in the sky is measured in a coarse grid for each redshift bin. The generated maps of neutral hydrogen (HI) flux are then used to trace the mass content of every pixel. These maps are then used to measure cosmology through their power spectrum as in e.g. \cite{ Abdalla:2004ah, Abdalla:2009wr}. The wide field of view (FoV) together with a lack of resolution makes it relatively cheap and quick to observe a large fraction of the sky. Intensity mapping can naturally not recover very small scale structure in the power spectrum, but with use of an appropriate FoV the Baryon Acoustic Oscillations (BAO) imprinted in the matter distribution can be recovered, as shown in \cite{Wyithe:2007rq}. 

An important issue when dealing with intensity mapping data is the presence of strong Galactic foregrounds. Radio emission from the Milky Way has an intensity up to five orders of magnitude higher than the HI signal we are interested in. It is therefore of great importance to develop sophisticated foreground removal techniques which do not leave considerable traces in the recovered signal. By analysing the cleaned maps we can measure to what degree the residuals of the Galactic foregrounds leftover in the data affect the cosmological information. For example, the power spectrum of the matter density distribution can be distorted by foreground removal contamination, which can cause systematic errors in the cosmological analysis.

There have been extensive studies of the Galactic foregrounds and foreground removal (see e.g. \citealt{ DiMatteo:2001gg, Oh:2003jy, 2004MNRAS.355.1053D, Santos:2004ju, Morales:2005qk, Wang:2005zj, 2008MNRAS.391..383G, Jelic:2008jg, 2009MNRAS.397.1138H, 2009MNRAS.398..401L, 2010A&A...522A..67B, 2011PhRvD..83j3006L, 
2011MNRAS.413.2103P, Chapman:2012yj, 2012MNRAS.419.3491L, 2013PhRvD..87d3005D, 2013ApJ...769..154M}) for the epoch of reionization data as well as the CMB. However, in the case of intensity mapping of the large-scale structure distribution at low redshifts, the field is relatively unexplored (\citealt{Ansari:2011bv}). 
We investigate how well a more realistic simulation of the foregrounds can be cleaned with the fast independent component analysis (\textsc{fastica}) method developed by \cite{DBLP:journals/tnn/Hyvarinen99} for intensity mapping experiments with future radio telescopes.

In this work, we use an SKA-like simulation of the HI distribution and combine it with a Galactic foreground simulation in the redshift range 0.01 to 1.5. We remove the foregrounds with the \textsc{fastica} method and evaluate its performance as a function of sky masks and other settings related to the \textsc{fastica} technique. The power spectrum estimations from the resulting intensity maps are used to evaluate the effect of the foreground subtraction on the cosmological analysis, where we specifically investigate the bias in the cosmological parameters and the recovery of the BAO scale. The systematic errors of the foreground removal change the broadband power of the cosmological signal, however they do not introduce a preferred scale. This motivates that cosmic distances measurements, in this case the BAO scale, are more robust than the power spectrum measurement. In this way, we create an end-to-end simulation of a future intensity mapping experiment, from input noisy data cube to cosmological parameter fits.

The paper is structured as follows. Sec.~\ref{sec-im} outlines the intensity mapping idea and reviews briefly the current state-of-the-art. We proceed with a detailed description of our simulation in Sec.~\ref{sec-data}. In Sec.~\ref{sec-cl}, the power spectrum estimation of the intensity maps and the theoretical modeling of the power spectrum is presented. We briefly describe the independent component analysis and its application to our data. The results of the foreground removed data are shown in Sec.~\ref{sec-fastica}. The cosmological parameter analysis and the resulting bias is presented in Sec.~\ref{sec-chi2}. We conclude in Sec.~\ref{sec-con} with the discussion of the impact of our results for future intensity mapping surveys.
\section{Intensity Mapping}
\label{sec-im}
The basic concept of an intensity mapping survey is mapping the integrated line flux of a voxel rather than measuring every single galaxy with exact redshift information. The advantage is that a large sky coverage is feasible within a relatively short observing time at the expense of low resolution. In order to recover the BAO scale, we chose the angle $\theta_{\rm{fwhm}}$ of the Gaussian beam which approximates the primary beam of the telescope, as 0.3 deg. At the redshifts of interest for our study $z \approx 0.7$, the BAO scale subtends an angle of order of $3 \deg$.

In principle, intensity mapping surveys are possible with every spectral line, for instance Lyman alpha in the optical (see \citealt{Peterson:2012hb, 2013arXiv1309.2295P}) or the rotational CO lines with e.g. $\nu_{1-0}=115\rm{GHz}$ (e.g. \citealt{2011ApJ...741...70L, Visbal:2011ee}) in the radio frequency regime.  However, the HI line with frequency $\nu=1.42\rm{GHz}$ is the commonly chosen line. Line confusion with different spectral line occurs if $\frac{\nu_1}{1+z_1}=\frac{\nu_2}{1+z_2}$, which often happens for different CO lines and also for Lyman alpha with other lines of the Lyman series. This kind of confusion is very insignificant \citep{2011ApJ...740L..20G} for the 21cm line because there is no other dominant spectral line close to its emitted frequency.

The ultimate intensity mapping experiment will be possible with the SKA, which is currently being planned. Also, there are SKA pathfinders, like ASKAP\footnote{\texttt{http://www.atnf.csiro.au/projects/mira/}} and MeerKAT\footnote{\texttt{http://www.ska.ac.za/meerkat/}}, which may be able to undertake an intensity mapping survey within a shorter timescale.

However, an HI intensity mapping survey can also be realised with a single, large $\approx 100$m dish, as shown with the Green Bank Telescope \citep{2013ApJ...763L..20M, 2013MNRAS.434L..46S}. There are other single-dish radio telescopes proposed such as the BINGO experiment (\citealt{Battye:2012fd}). BINGO will observe the HI emission between redshifts $0.13<z<0.48$ over $2000\text{deg}^2$ in 1 year of observing time.
A slightly different approach is chosen by the CHIME\footnote{\texttt{http://chime.phas.ubc.ca/ }} design which uses cylindrical dishes as elements of an interferometer. The aim of the experiment is measuring the HI flux in a volume of 300 $\text{Gpc}^3$ covering the redshift range $0.8<z<2.5$. Another more recent planned project is the Tianlai\footnote{\texttt{http://tianlai.bao.ac.cn/}} project (\citealt{Chen:2012xu}).

These planned surveys give very promising prospects for the future of intensity mapping. They will cover a very wide redshift range as well as large fraction of the observable sky.
\section{Simulated Data}
\label{sec-data}
In this section, we describe the data simulations used in this analysis. These simulations can be split in four subcategories: the cosmological signal based on the SKA design study (SKADS) simulation \citep{Wilman:2008ew}, the Galactic foreground \citep{Shaw2013}, the noise estimator and the lognormal realisations used for the computation of the covariances.
\subsection{Wilman SKA Simulation}
\label{1sec:wilman}
The large-scale matter distribution used in this work is a semi-empirical simulation of the radio continuum sky up to redshift 20 as described in \cite{Wilman:2008ew}. A brief description of the properties of the simulation is given below. For a detailed description we refer the reader to \cite{Wilman:2008ew}\footnote{\texttt{http://s-cubed.physics.ox.ac.uk/s3\_sex}}. 

The simulation is based on a realisation of the linear matter power spectrum produced by CAMB (\citealt{Lewis:1999bs}). The cosmological model used in the Wilman SKA simulation is: 
$\Omega_{\rm{m}}=0.3$, $\Omega_{\rm{k}}=0.0$, $w=-1.0$, $h=0.7$, $f_{\rm{baryon}}=0.16$, $\sigma_8=0.74$, $b=1.0$ and $f_{\rm{NL}}=0$. This density field realisation is gridded in cells of size $5\rm{Mpc}/h$ from which galaxies are sampled. 
The galaxy bias function $b(z)$ follows the description of \cite{Mo:1995cs} with a cut-off redshift for different galaxy types. This cut-off is chosen so that the bias is held constant above a give redshift to prevent exponential blow-up of the clustering. Galaxy clusters are identified by looking for regions with overdensities larger than the critical density with use of the Press-Schechter (\citealt{Press:1973iz}) and Sheth-Tormen formulations (\citealt{Sheth:1999mn}). The cell design of the simulation leads to a quantisation of the cluster masses.
The  simulation includes populations of four types of galaxies: radio-quiet active galaxy nuclei (AGN) (\citealt{Jarvis:2004gh}), radio-loud AGNs (\citealt{Willott:2000dh}) of low and high luminosities and star-forming galaxies (\citealt{Yun:2001jx}). The empirical luminosity functions of the different sorts of galaxies are extrapolated to high redshifts, since there are no relevant observations available in this regime so far. The HI masses of the galaxies are correlated with the star formation rate of galaxies \citep{Wilman:2008ew}. They are assigned according to the correlation between the $1.4\rm{GHz}$ luminosity function given by \cite{2001ApJ...558...72S} and the HI masses \citep{2005MNRAS.361...34D}. This description is only valid for star forming galaxies and, due to the lack of an irregular galaxy population in the simulation, the resulting HI mass function does not exactly match the locally observed mass function by \cite{Zwaan:2003hp} in the redshift range $0<z<0.1$. 

In this work, we made use of a half sky extension of the publicly available 20x20 degree simulation and processed a half-sky simulation. The half-sky simulation has some limitations compared to the previously described one: it does not include AGNs and clusters of galaxies and only extends to redshift 1.5. The initial resolution of the cells was also reduced by a factor of two. For the purpose of our analysis, these changes are not significant, since intensity mapping uses coarse resolution and we want to examine the potential of future radio observations in a low redshift regime. Also, we are most interested in the star forming objects in this paper. 

There are other, more accurate approaches to simulating galaxy distributions, such as hydrodynamical or semi-analytic methods. In this study, we examine intensity maps with half-sky coverage to high redshifts which cover a very large volume not currently available with other simulation approaches. In addition, the semi-empirical simulations are by construction tuned to reproduce the observational properties of the underlying galaxy distribution.

We constructed maps of the brightness temperature of the HI emission using these simulations. We converted the galaxy catalogue of the Wilman simulation into 21-centimeter brightness temperature maps following the description in \cite{Abdalla:2009wr}. Given the HI mass $M_{\rm{HI}}$, the neutral hydrogen emissivity per steradian can be calculated using the emission coefficient of the 21cm line transmissions $A_{12}$. We used the Rayleigh-Jeans law for low frequencies to convert the emissivity into the measurable brightness temperature per pixel
\begin{equation}
T=\frac{3 A_{12} h c^2}{32\pi m_{\rm H} k} \frac{M_{\rm{HI}} }{\chi^2(z) \Delta \nu \nu_{21} \Omega_{\rm{pix}} }
\end{equation}
where  $h$ is the Planck constant, $m_{\rm H}$ is the mass of the hydrogen atom and $k$ is the Boltzmann constant.
In addition we use the comoving distance $\chi(z)$, width of the frequency slice $\Delta \nu$ and the solid angle of one resolution element $\Omega_{\rm{pix}}$. All the maps shown in this paper are in units of Kelvin.

Following the temperature conversion, the maps are smoothed to the resolution of the intensity mapping survey. We assume a Gaussian primary beam with a FoV constant in redshift. The solid angle of each resolution element is calculated via $\Omega=1.33\theta_{\rm{FWHM}}^2$, where $\theta_{\rm{FWHM}}$ is the opening angle at the full width half maximum (FWHM) which is chosen as $0.3\deg$ throughout this study.

\begin{figure*}
\begin{tabular}{ccc}             
                  \includegraphics[width=0.8\textwidth, angle=0, clip=true, trim= 0 210 0 320]{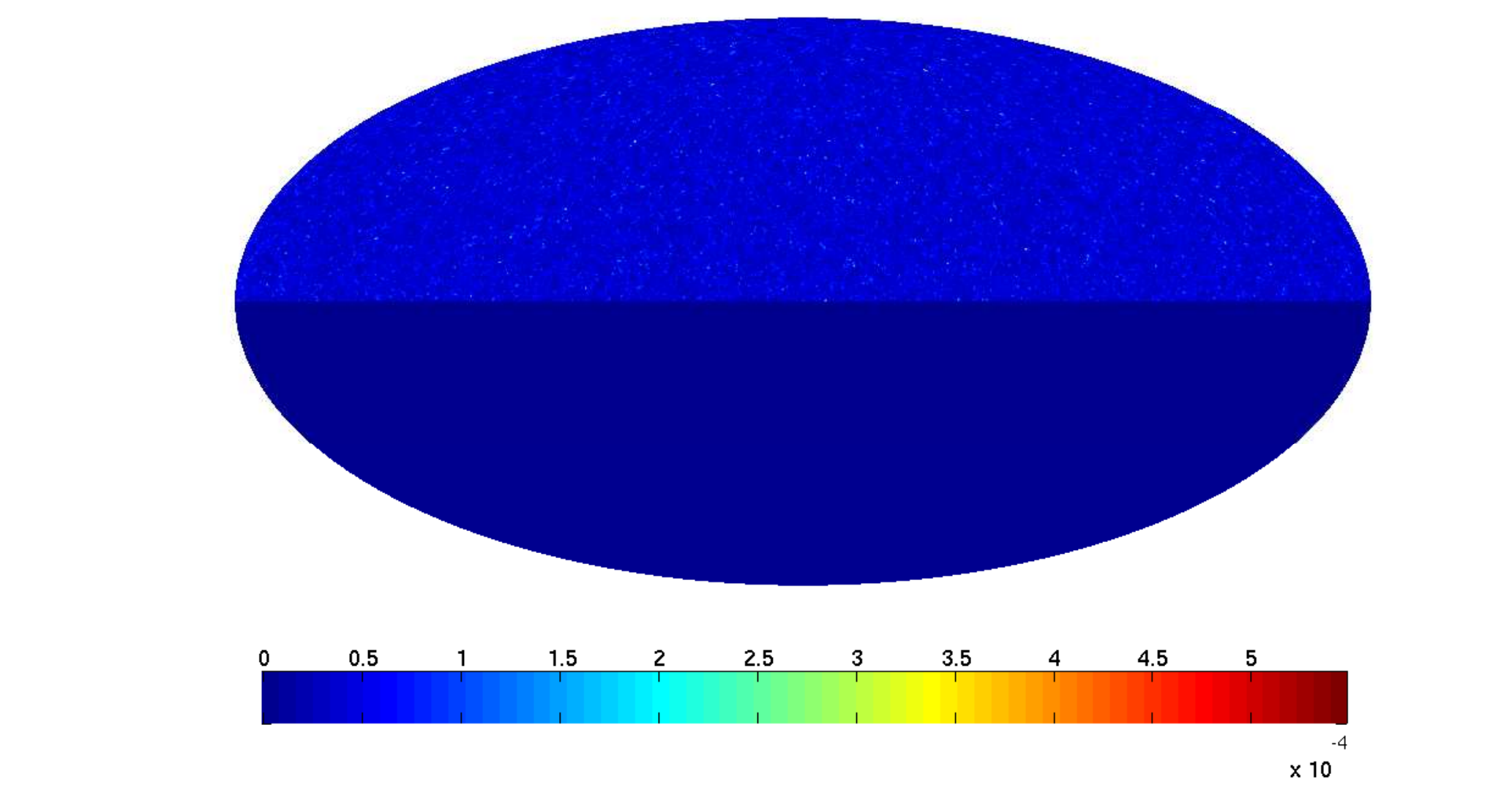}\\

                 \includegraphics[width=0.8\textwidth, angle=0, clip=true, trim= 0 210 0 320]{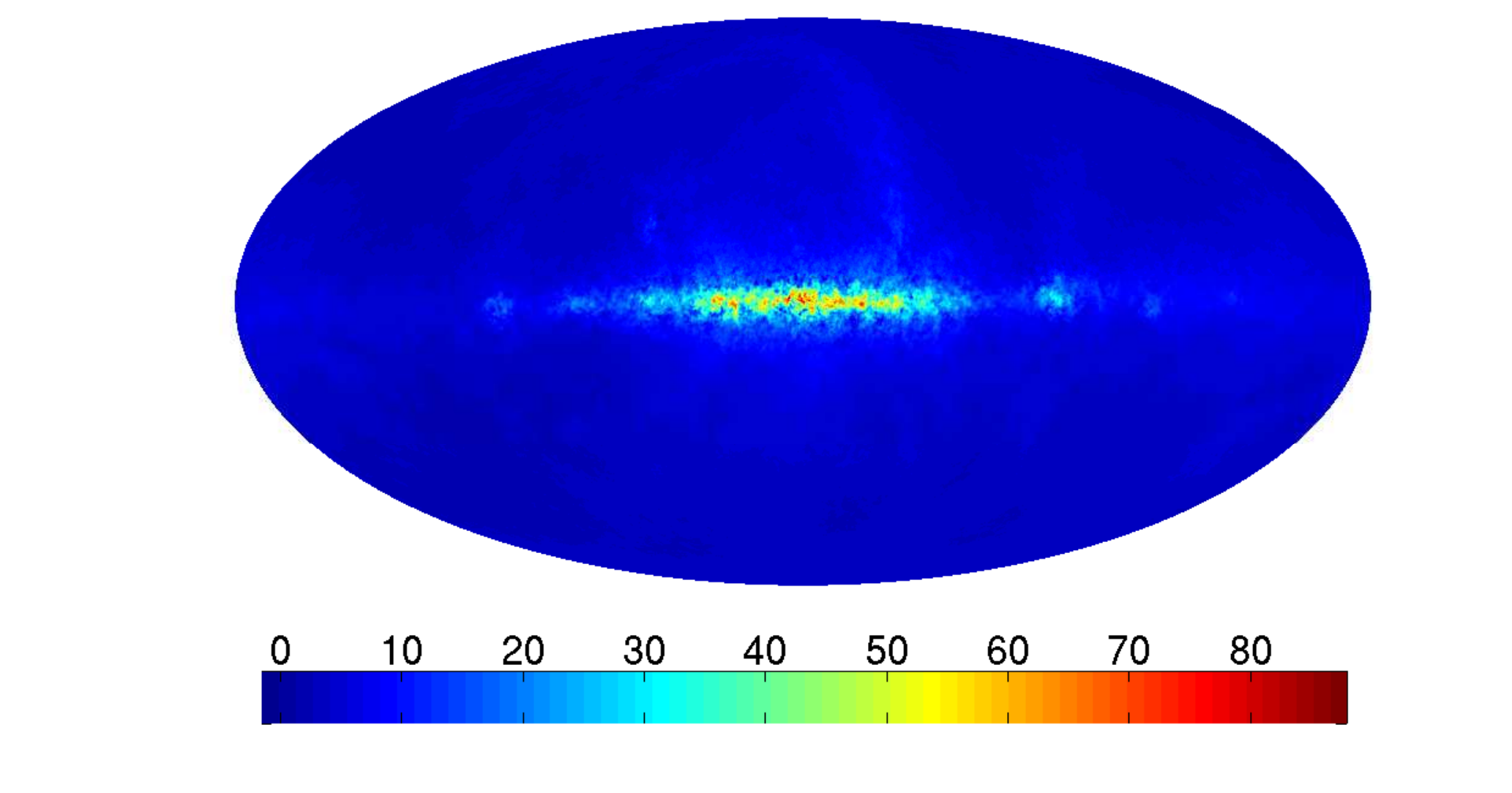}\\
                                                         
         \includegraphics[width=0.8\textwidth, angle=0, clip=true, trim= 0 210 0 320]{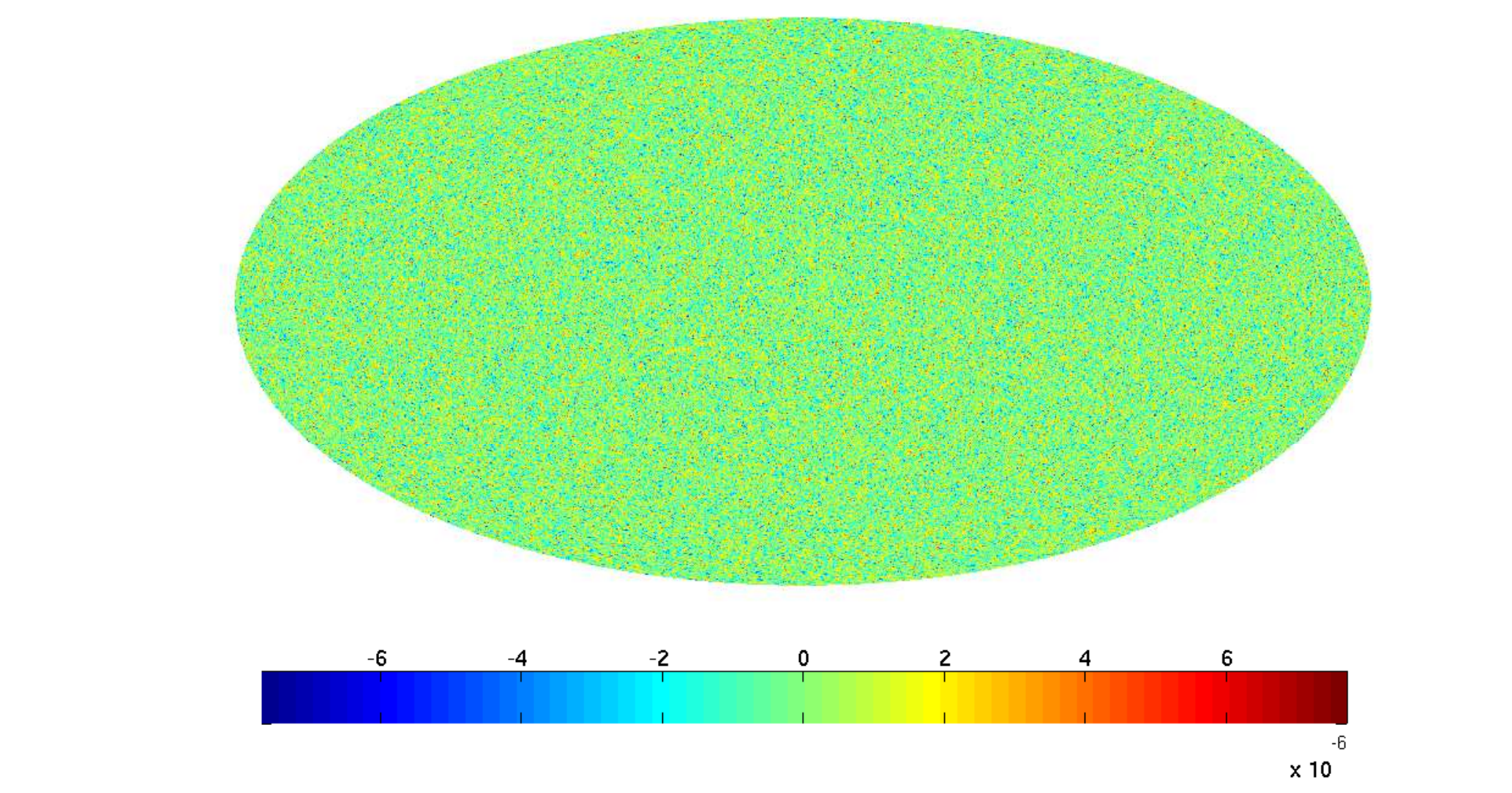}\\
\end{tabular}
\caption[Simulated Data, Galactic Foregrounds and Noise Maps]{ Simulated Data (upper panel),Galactic Foregrounds (middle panel)and Noise Maps (lower panel) for frequency $ \nu=829\rm{MHz} $ with frequency width $\Delta \nu=5.2\rm{MHz}$ which corresponds to the redshift slice $0.71<z<0.72$. The colour bar represents brightness temperature in units of K.}
\label{fig:maps}
\end{figure*}

In Fig.~\ref{fig:maps} in the first panel, one example map of the Wilman simulation with frequency width $\Delta \nu=5.2\rm{MHz}$ is pictured. This narrow binning width is used in the Galactic foreground removal in order to improve the performance. In the cosmological analysis wider binning is used so that we do not lose the correlations due to BAO and compress the data enough to obtain a reasonably-sized dataset.
\subsection{Galactic Foreground}
\label{1sec:galfg}
The main contributions to the radio continuum emission of galaxies at the frequencies of interest are synchrotron and free-free electron emission (\citealt{Condon:1992rq}). The foregrounds produced by our Galaxy are well-studied in the frequency range of the microwave background emission above 10GHz. Our frequencies of interest ($570<\nu<1400 \rm{MHz}$) are less explored and there are only two all-sky maps at $\nu=408\rm{MHz}$ (the Haslam map by \citealt{Haslam:1982zz}) and $\nu=1420\rm{MHz}$ (\citealt{Testori:2001vp}). 

The foreground simulation used here is described in detail by \cite{Shaw2013}. In this simulation the Haslam map is extrapolated with the help of a sky map of the radio spectral index, since observations have shown that the spectral index of the foreground is not constant with sky latitude. These maps of the spectral index are derived using the global sky model by \cite{deOliveiraCosta:2008pb}. This method performs a principal component analysis (PCA) of the most relevant observations of the Milky Way in the radio regime and determines that the first three PCs are sufficient to create a global sky model of our galaxy. In addition, it takes the variation of the spectral index with latitude on the sky and frequency into account. 
We improve the resolution of our foreground maps, beyond that of the Haslam,
by simulating small scale variations. These are drawn from a multi-frequency
angular power spectrum constructed following the description in
\cite{Santos:2004ju}. This uses power laws to model the angular fluctuations and
frequency auto-correlations, but an exponential decorrelation between
frequencies. Any large-scale variations in this realisation, which are already
included in the Haslam map, are removed and, to assure the smoothness of the
total power spectrum, the variance on smaller scales than five degrees is
rescaled to the fluctuations of the Haslam map. This gives frequency structure
beyond a pure power law, and adds in arbitrarily small angular structure,
while remaining consistent with observational data. A Galactic foreground map
of our simulations is shown in the second row of Fig.~\ref{fig:maps}.
\subsection{Lognormal Realisations}
\label{1sec:lognormal}

In our analysis, we use the power spectrum of the large-scale structure in redshift shells to quantify the statistical properties of our simulated dataset. To determine the covariance between data points in our simulated sky accurately, we require many mock realisations. Unfortunately, we only have one simulated SKA sky to analyse. We therefore built 100 lognormal realisations of the matter distribution from the same power spectrum used in the SKA simulation with the same redshift distribution of the HI-selected galaxies. We extracted their power spectra and cross-correlated them in order to estimate the covariance matrices of the power spectra of the simulated data between different angular scales and redshifts. 

In order to compute the covariances in the power spectra of an intensity mapping observation, the lognormal realisation need to be converted into temperature maps of the 21cm flux.
The HI masses of the galaxies of the lognormal simulations are assigned from the HI mass distribution according to the properties present in the SKA simulation. Specifically, we compute the cumulative histogram of the HI mass distribution in each redshift bin of the simulation. We invert the cumulative histogram of the HI mass and fit a spline function to it. We draw a random number between 0 and 1 for each galaxy, multiply it by the total galaxy number and use the fitted spline function to draw the HI mass of each galaxy. With this procedure, the total HI density is preserved for each redshift bin. In the subsequent analysis, the lognormal galaxy catalogues  are converted into smoothed temperature maps as described in Sec. \ref{1sec:wilman}.
\subsection{Receiver Noise}
\label{1sec:noise}
The receiver noise of a radio interferometer (\citealt{thompson04, Abdalla:2009wr}) in the measured brightness temperature is given by
\begin{equation}
\Delta T_{\rm{b}}=\frac{c^2 T_{\rm{Sys}}}{\nu^2 \Omega f A_{\rm{eff}}\sqrt{2\Delta \nu t} }.
\label{equ:rmst}
\end{equation}
$T_{\rm{Sys}}$ describes the system temperature of the radio receivers which has a goal of 10K for the full SKA design, $t$ is the integration time and $A_{\rm{eff}}$ is the effective collecting area of the telescope. It is convenient to combine the system temperature and the collecting area into one quantity since increasing the area has the same effect as decreasing the system temperature. The SKA design goal (\citealt{Carilli:2004nx}) is $A_{\rm{eff}}/T_{\rm{Sys}}=2.0\cdot 10^4 \rm{m}^2\rm{K}^{-1}$. We are assuming a survey duration of 6 months, which in a case of a half-sky observation for our FoV results in an integration time per pointing of 77s.

The parameter $f$ describes what percentage of the full SKA configuration is being simulated. $f$ ranges between 0 and 1.
The rms of the brightness temperature scales as $1/\sqrt{N}$ with the number of pixel feeds $N$. We simulate the instrumental noise according to a ten percent SKA realisation with $N=25$ pixel feeds. However, $f$ and $N$ are completely degenerate, so this noise level would also describe a 50\% SKA observation with a single pixel feed. There are several other instrumental setups which can produce similar noise properties. For instance, single dishes, rather than interferometers may be used to produce intensity mapping experiments. We do not discuss the advantages or disadvantages of such approaches in this paper.

As introduced in Sec. \ref{1sec:wilman}, $\Omega=1.133\theta_{\rm{FWHM}}^2$ is the solid angle of our assumed Gaussian primary beam with opening angle $\theta_{\rm{FWHM}}$ . 
We simulate a map of Gaussian noise smoothed to the width $\theta_{\rm{FWHM}}=0.3\deg$. For the foreground removal, we chose a binning of the maps constant in frequency with bin width $\Delta \nu= 5.2\rm{MHz}$. A noise map according to the chosen frequency binning of the SKA simulation is pictured in Fig.~ \ref{fig:maps} in the third row. 
\section{Power Spectrum Estimation}
\label{sec-cl}
\subsection{Data Measurement}
Assuming an observation in the redshift bin $z_i$, it is convenient to expand the matter distribution in spherical harmonic functions since we are observing the galaxy distribution on a sphere.
\begin{equation}
\sigma(\theta, \phi)= \sum_{\ell =0}^{\infty} \sum_{m=-\ell }^{\ell } a_{\ell m} Y_{\ell m}(\theta, \phi)
\label{Sphericalharmonic}
\end{equation}
where $\sigma$ is the surface density of galaxies at a given direction in the sky, with coefficients $a_{\ell m}$ and spherical harmonics $Y_{\ell m}(\theta, \phi)= \sqrt{\frac{(2\ell +1)(\ell -m)!}{4\pi(\ell +m)!}}P^m_\ell (\cos\theta)e^{im\phi}$. Here $P^m_\ell $ denote the Legendre polynomials. The spherical harmonics are normalised such that $\int  Y_{\ell m}(\theta, \phi)Y_{\ell 'm'}^*(\theta, \phi) d\Omega=\delta_{\ell \ell '}\delta_{mm'}$. Multiplying Eq~ \ref{Sphericalharmonic} by $Y_{\ell 'm'}^*(\theta, \phi)$ and integrating over the solid angle $\Omega$ defines the back transformation 
\begin{equation}
a_{\ell m}=\int  \sigma(\theta, \phi) Y_{\ell m}^*(\theta, \phi) d\Omega.
\label{sphericalcoeff}
\end{equation}
The coefficients $a_{lm}$ completely describe the properties of the galaxy distribution in spherical harmonic space. For an intensity mapping survey, the map consists of discrete temperature values $T$ for every resolution element with solid angle $\Delta \Omega_{\rm{pix}}$ rather than a continuous galaxy distribution $\sigma$. Hence, the integral in \ref{sphericalcoeff} becomes a summation over the pixels $p_i$ of the map.
\begin{equation}
a^{\rm{pix}}_{\ell m}=\sum_{i=1}^{N_{\rm{pix}}} T(p_i)Y_{\ell m}^*(p_i) \Delta \Omega_{\rm{pix}}
\label{sphericalcoeff_pix}
\end{equation}
The power spectrum $C(\ell )$ is calculated via the autocorrelation of the spherical harmonic expansion. It contains information about the preferred correlation length of matter overdensities in the distribution. If the matter distribution is a realisation of a Gaussian random field, the statistical properties of the distribution are completely described by the power spectrum and the mean of the expansion coefficients vanishes: $\left<a_{\ell m}\right>=0$. Then the variance is $\left<a_{\ell m}a_{\ell m}^*\right>=C(\ell )$, which leads to the estimator of the power spectrum in spherical harmonic space
\begin{equation}
C(\ell )=\frac{1}{2\ell +1}\sum_{m=-\ell }^{\ell }|a_{\ell m}|^2.
\label{Cldefinition}
\end{equation}

In most cases, there is no full sky observational data available. In terms of power spectrum estimation, this means that the estimator of Equ.~\ref{Cldefinition} is biased, and measurements at different multipoles are correlated. In order to compare cut-sky observations to theoretical predicted full-sky power spectra, we need an estimator for the full sky power spectra. In this work, we chose to account for the correlation and loss of power that the cut sky induces with the Peebles approximation (\citealt{1973ApJ...185..413P}). However, there are also other approaches to correct for this, most commonly the maximum likelihood approach (e.g. \citealt{Efstathiou:2003tv}).
\begin{equation}
C_{\rm Peebles}(\ell )=\frac{1}{2\ell +1}\sum_{m=-\ell }^{\ell }\frac{|a_{\ell m}^{\rm gal}-\frac{N}{\Delta\Omega}I_{\ell m}|^2}{J_{\ell m}}
\label{Peebles}
\end{equation}
with
\begin{equation}
I_{\ell m}=\int_{\Delta\Omega}  Y_{\ell m}^*(\theta, \phi) d\Omega; \hspace{5pt} 
J_{\ell m}=\int_{\Delta\Omega}  |Y_{\ell m}(\theta, \phi)|^2 d\Omega.
\label{eq:IlmJlm}
\end{equation}
Comparing $I_{\ell m}$ to Equ.~\ref{sphericalcoeff}, it represents the spherical harmonic coefficients of the mask, which has a value of 1 for areas within the observation and zero elsewhere. For a full sky observation $I_{\ell m}$ should be zero in all components. In addition, $J_{\ell m}$ weights the power spectrum according to the sky fraction such that it is 1 for complete sky. For a discrete analysis of intensity maps the integrals of Eq.~ \ref{eq:IlmJlm} become sums over all pixels of the map.

\subsection{Theoretical Prediction}
\label{2sec:theory}
A theoretical prediction for the matter power spectrum $P(k,z)$ is obtained using a present day linear power spectrum estimate $P(k, z=0)$ by CAMB (\citealt{Lewis:1999bs}). 
The evolution of the power spectrum in time is estimated following linear perturbation theory in the Newtonian description. The resulting differential equation for the growth factor reads
\begin{equation}
\ddot D(t)+\frac{2\dot a}{a}\dot D(t)=4\pi G \bar \rho D(t)
\end{equation}
The resulting approximation for the time-dependent power spectrum is
\begin{equation}
P(k,z)=D^2(z)P(k, z=0)
\end{equation}
The projected angular power spectrum $C(\ell )$ in linear theory can be calculated via \cite{Blake:2004pb}
\begin{equation}
C(\ell )=\int P_0(k)W(\ell ,k) dk
\end{equation}
where $W(\ell ,k)$ is a window function, which projects the power spectrum onto the sphere. For a more elaborate discussion see \cite{Huterer:2000uj, Tegmark:2001jh}. Since our particular interest is not on the very large-scales of the matter distribution, we use the small angle approximation for computational ease.
\begin{equation}
C(\ell )=b^2\int P\left(k=\frac \ell {\chi(z)}, z\right)\chi^{-2}(z)p^2(z)\left(\frac{d\chi(z)}{dz}\right)^{-1}dz
\end{equation}
which is valid for large $l$ (\citealt{Blake:2004pb, Blake:2006kv}). 

The bias function $b$ reflects the fact that we observe the visible galaxy power spectrum, whereas the above derivations are valid for the matter distribution. Therefore we introduce the linear bias as $P_{\rm{gal}}(k,z)=b^2(z)P(k,z)$. In the following, we will assume a redshift independence of the bias $b(z_i)=b$ within the redshift shell and marginalize over the unknown bias in the following analysis. The probability distribution of the galaxies $p(z)$ is set to constant within every redshift shell such that $p(z)=1/\Delta z$. In this work, we neglect redshift space distortions since they only influence the large-scales of the power spectrum. 

\begin{figure}
\includegraphics[width=0.5\textwidth, clip=true, trim= 0 0 0 10]{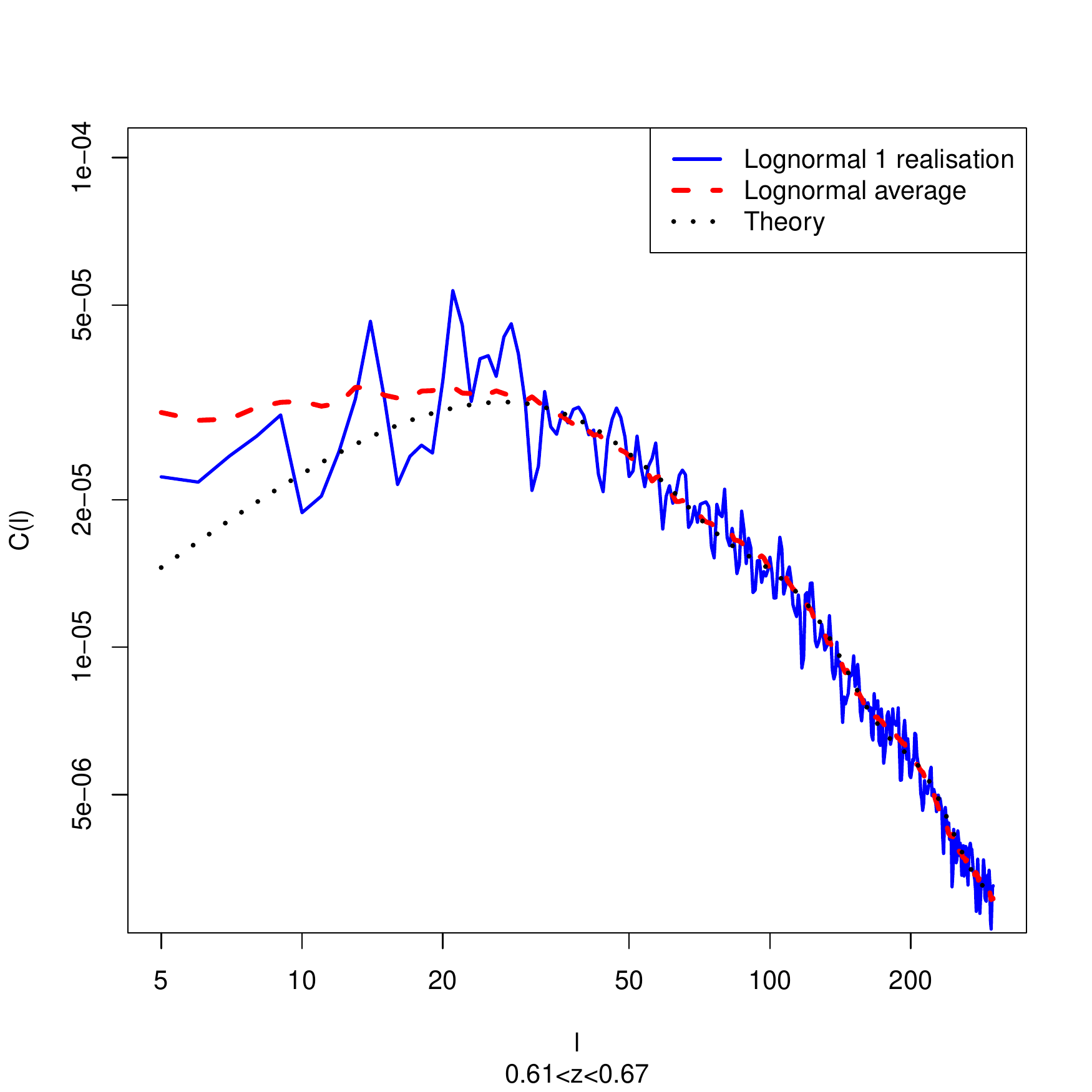}
\caption{Theoretical prediction for the large-scale structure power spectrum (dotted black line) in comparison with estimated power spectrum of one lognormal realisation (solid blue line) and the average of the estimated power spectrum of 100 lognormal realisations (dashed red line). }
\label{fig:theorcl}
\end{figure}
In Fig.~ \ref{fig:theorcl} the theoretical prediction of the $C(\ell )$ for redshift $0.61<z<0.67$ is pictured as the dotted black lines. 
The temperature power spectrum of one lognormal realisation (blue line) and the average lognormal $C(\ell )$ (red line) are plotted. On large-scales they mismatch the black line of the theoretically modeled power spectra due to the use of the small angle approximation, however, the agreement is good on scales $l>20$. The graphs are shown with the shot noise removed. 

\subsection{Noise estimate}

The noise present in the power spectrum measurement can be estimated as
\begin{equation}
\sigma_{\rm{noise}}(C(\ell ))=\sqrt{\frac{2}{f_{\rm sky}(2\ell +1)}} \left( C(\ell )+ \it{N}(\ell )\right).
\label{noiseeq}
\end{equation}
The first term in Eq.~ \ref{noiseeq} refers to the cosmic variance which describes the fact that we can only observe one realisation of the Universe. The second term $\it{N}(\ell )$ is the signal expected for an unclustered distribution, which corresponds to shot noise in the case of galaxy distributions, or other sources of noise such as telescope noise in the case of intensity mapping surveys.


The shot noise of the galaxy power spectrum is equal to the solid angle of the survey area divided by the total number of observed galaxies. For the temperature maps, we had to simulate the shot noise by creating a uniform distribution of galaxies in the absence of clustering with the same total galaxy number as the input simulation. We then draw HI masses from the mass distribution of the SKA simulations and converted the galaxy maps into brightness temperature maps. The estimated power spectrum of the temperature noise maps is an estimate of the temperature shot noise of the SKA simulations. This procedure gave the estimate of the shot noise contribution to $\it{N}(\ell )$.
 
The beam of an intensity mapping survey is rather wide, in this study we approximated it with a Gaussian beam. We realised this through smoothing the input maps of cosmic signal, beam noise and Galactic foreground to a $\theta_{\rm{FWHM}}=0.3\deg$. The Gaussian beam implies that small scales of the power spectra are suppressed. In order to recover the original power spectrum we deconvolve it using the beam window function of a Gaussian beam
 \begin{equation}
 W_{\rm{beam}}(\ell )=\exp{-\frac{\ell (\ell +1)}{2\ell_{\rm{beam}}^2}}
 \label{eq:beamwindow}
 \end{equation}
 where $\ell _{\rm{beam}}=\frac{\sqrt{8\log(2)}}{\theta_{\rm{FWHM}}}$. 
 With Eq.~ \ref{eq:beamwindow} we can also estimate the receiver noise power spectrum as
 \begin{equation}
 C_{\rm{noise}}(\ell )=\sigma_{\rm{rms}}^2 \Omega_{\rm{pix}} W_{\rm{beam}}(\ell )
 \label{eq:shotnoise}
 \end{equation}
where the root-mean-square (rms) $\sigma_{\rm{rms}}$ is the temperature rms of the unsmoothed map given by Eq.~\ref{equ:rmst}. This theoretical description for the receiver noise gives the second contribution to $\it{N(l)}$, which we add to the shot noise term.
\subsection{Lognormal Covariance}
\label{sec:cov}
As described in section \ref{1sec:lognormal}, we estimated the covariance matrices in multipole and redshift space with $N_{\rm{real}}=100$ realisations of a lognormal galaxy distribution. The covariance in the measured power spectrum between two multipoles $\ell$ and $\ell '$ for one redshift bin $z_i$ is calculated via 
\begin{equation}
\rm{Cov}_{z_i}(\ell ,\ell ')=\frac{1}{N_{\rm{real}}-1}\sum_{p=1}^{N_{\rm{real}}} (C_{p,z_i}(\ell )-\bar C_{z_i}(\ell ))(C_{p,z_i}(\ell ')-\bar C_{z_i}(\ell '))
\end{equation}
where  $C_{p,z_i}(\ell )$ is the power spectrum of the $p\rm{th}$ realisation and $\bar C_{z_i}(\ell )$ is the average power spectrum of the $N_{\rm{real}}$ realisations in one redshift bin $z_i$. The covariance between identical multipoles at different redshifts is similarly defined as 
\begin{equation}
\rm{Cov}(z_i,z_j)=\frac{1}{N_{\rm{real}}-1}\sum_{p=1}^{N_{\rm{real}}} (C_{p,z_i}(\ell )-\bar C_{z_i}(\ell ))(C_{p,z_j}(\ell )-\bar C_{z_j}(\ell )).
\end{equation}

In the following plots, we show the correlation matrix instead of the covariance matrix, where the diagonal is normalised to one via
\begin{equation}
\rm{Cor}_{z_i}(\ell ,\ell ')=\frac{\rm{Cov}_{z_i}(\ell ,\ell ')} {(\sqrt{\rm{Cov}_{z_i}(\ell ,\ell )\rm{Cov}_{z_i}(\ell ',\ell ')}) }.
\end{equation}

\begin{figure*}
\centering
\subfigure[Original data]{         
                  \centering 
\includegraphics[width=0.5\textwidth, clip=true, trim= 0 28 0 20]{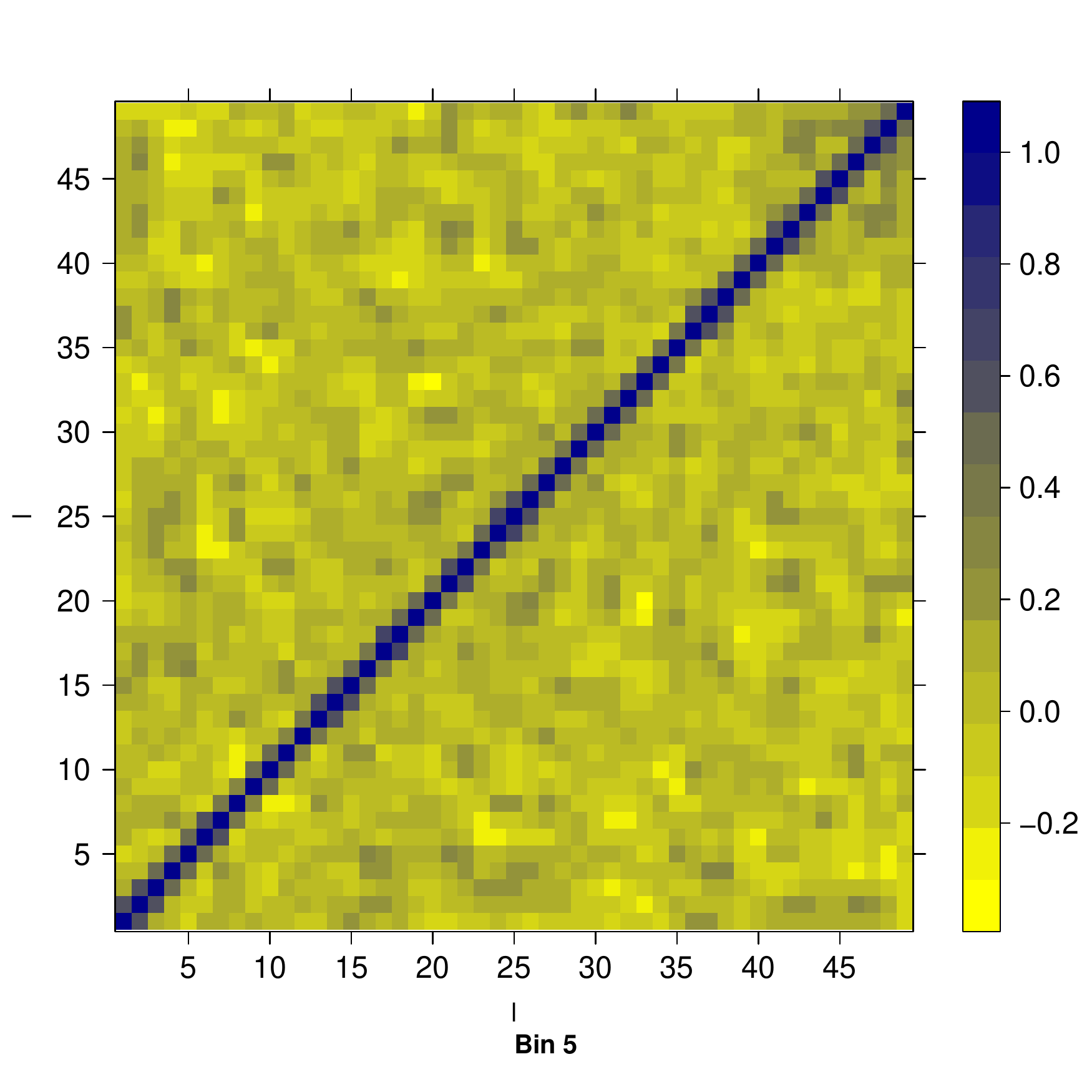} 
 \label{fig:corrmultipolepre}
        }%
        \subfigure[After Foreground Removal]{         
                  \centering  
\includegraphics[width=0.5\textwidth, clip=true, trim= 0 28 0 20]{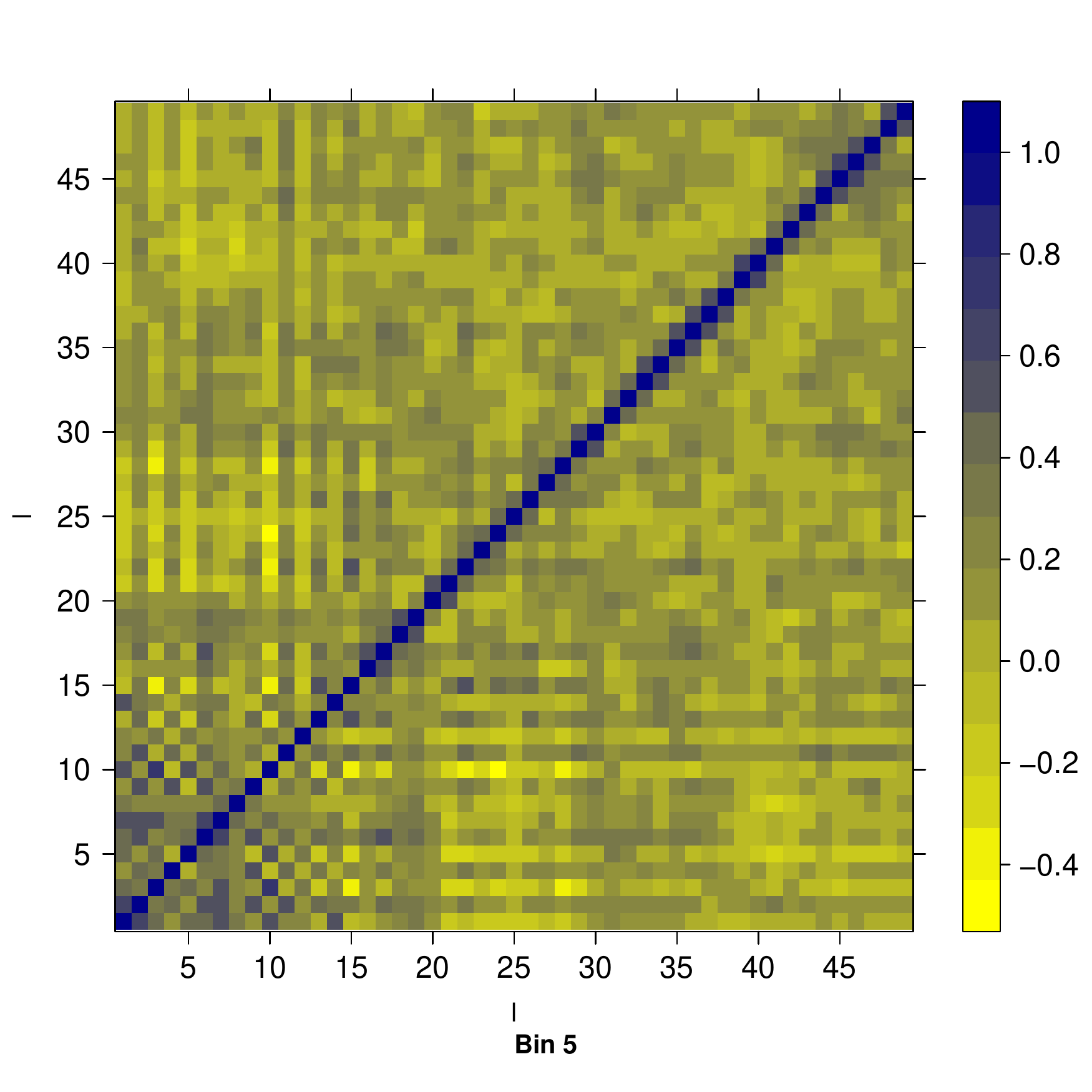}
 \label{fig:corrmultipolepost}
}
\caption{The correlation matrix between power spectrum measurements for the multipole range $1<\ell <50$ for an example redshift slice $1.11<z<1.16$, computed with the original lognormal realisations in the left panel and including the foreground removal systematics in the right panel. The high values of the elements in the secondary diagonals show the correlation between adjacent multipoles introduced by the sky cut. Additional covariance is introduced by foreground subtraction.}
\label{fig:cormultipole}
\end{figure*}
\begin{figure*}
\centering
\subfigure[Original data]{         
                  \centering  
			\includegraphics[width=0.5\textwidth, clip=true, trim= 0 28 0 20]{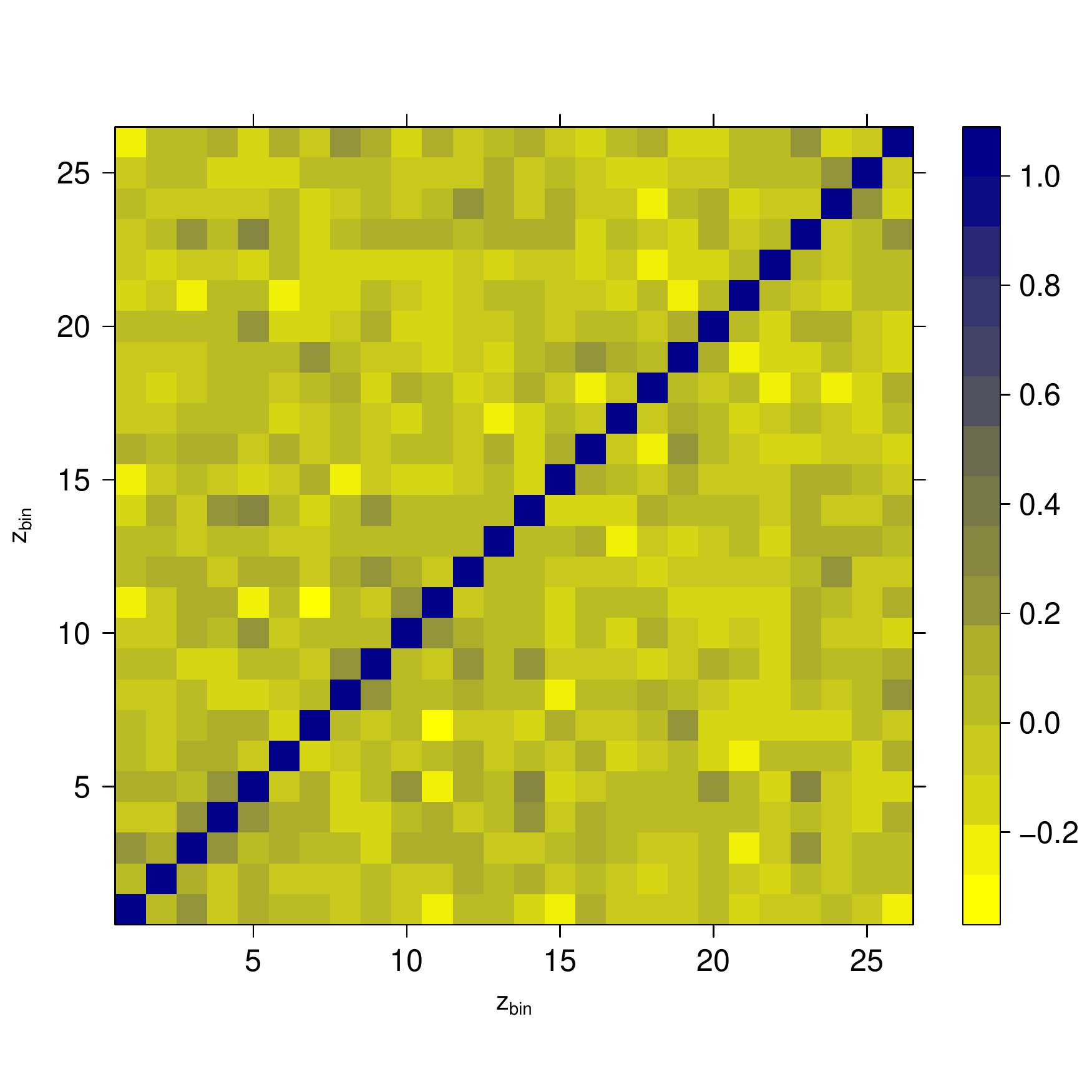} 
                \label{fig:corrredshiftpre}
        }%
        \subfigure[After Foreground Removal]{         
                  \centering  
	\includegraphics[width=0.5\textwidth, clip=true, trim= 0 28 0 20]{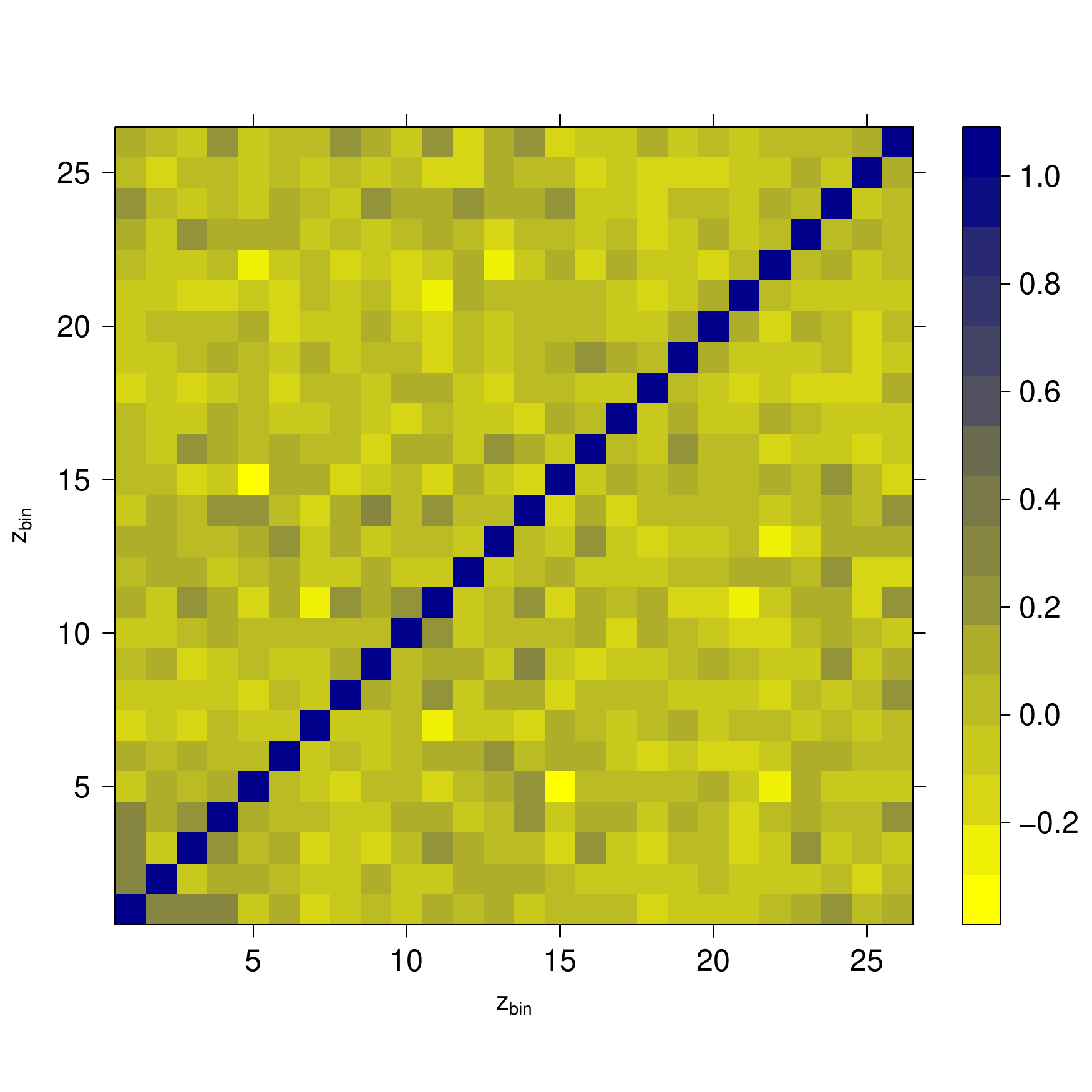}
                \label{fig:corrredshiftpost}
        }
\caption{The correlation matrix between power spectrum measurements at different redshifts for an example multipole $\ell =50$, computed with the original lognormal realisations in the left panel and including the foreground removal systematics in the right panel.}
\label{fig:corredshift}
\end{figure*}

Figures \ref{fig:cormultipole} and \ref{fig:corredshift} give an example of the correlation matrices in multipole for a given redshift bin or in redshift for a given multipole, respectively, showing results both before and after foreground removal. 
It can be seen in the left panel of Fig.~ \ref{fig:cormultipole} that the correlation between adjacent multipoles is relatively high. The $\rm{Cor}(\ell , \ell +1)$ is pictured as the secondary diagonal of the matrix. The other off-diagonal entries are very low as expected for a survey covering sky fraction 0.5. The right hand side shows the correlation matrix of the power spectra after the foreground removal. It significantly more contaminated with a complex structure due to correlations introduced by \textsc{fastica}.
In Fig.~ \ref{fig:corredshift}, the redshift correlations between the bins are shown for one multipole $\ell =50$. It can be seen in the left panel that the original power spectra has negligible correlations between different redshift slices. After the foreground removal on the right hand side, the correlations are still very small.

The binned power spectra are weighted according to 
\begin{equation}
C(\ell ')=\frac{\sum_{\ell =\ell '}^{\ell '+\delta \ell }(2\ell +1)C(\ell )}{\sum_{\ell =\ell '}^{\ell'+\delta \ell }(2\ell +1)}.
\end{equation}
Given that power spectrum measurements at adjacent multipoles exhibit high correlations in Fig.~ \ref{fig:cormultipole}, the power spectra are rebinned with $\delta \ell =2$ for the final analysis. In addition, these tests showed that relative to the diagonal, the entries of the correlation matrix can be assumed to be negligible. 
Therefore, for computational ease, we choose to approximate the covariance matrix in a diagonal form using the expression given in Eq.~ \ref{noiseeq} where we use the theoretical prediction as $C(\ell )$.
%
%
\section{Foreground Removal}
\label{sec-fastica}
In this section, the foreground removal technique and its applications to the data are shown. We introduce a masking technique to improve the performance and show the effects on the residuals of the reconstructed cosmological signal.

\subsection{\textsc{fastica} Technique}\label{3sec:fastica}
In the following subsection, the basic principles of the \textsc{fastica} method 
(\citealt{DBLP:journals/tnn/Hyvarinen99}) are outlined. For a comprehensive tutorial, we recommend this online tutorial\footnote{\texttt{http://cis.legacy.ics.tkk.fi/aapo/papers/IJCNN99\_tutorialweb/ }}
and for scientific applications refer for example to \cite{Maino:2001vz, Bottino:2009uc, Chapman:2012yj}.

The  measurement $\bmath x$ is considered to be a linear combination of independent components $\bmath s$. The basic equation of the independent component analysis (ICA) is 
\begin{equation}
\bmath x= \mathbf A \bmath s= \sum_{i=1}^{N_{\rm{IC}}}\bmath{a_i} s_i 
\label{eq:ica}
\end{equation}
where $\mathbf A$ is the mixing matrix which contains the weights of the single ICs of which the measured signal is expressed. The columns of the mixing matrix are referred to as $\bmath{a_i}$. The dimensions of $\mathbf A$ are (Number of independent components) $\times$ (Number of measurements). The inverse of Eq.~ \ref{eq:ica}, which is required to determine the unknown ICs of the measurements, is 
\begin{equation}
\bmath s= \mathbf W\bmath x
\label{eq:invica}
\end{equation}
where $\mathbf W$ is the weighting matrix which is defined as the inverse of $\mathbf A$. A fundamental assumption of this technique is that the ICs have to be statistically independent, which implies that their joint probability density functions (pdf) is the product of the single pdfs of the variables: $p(y_1,..,y_n)=\prod_{i=0}^n p_i(y_i)$.
This also implies that the expectation value of the joint functions $f_i(y_i)$ is: $E\{f_1(y_1),...,f_n(y_n)\}=\prod_{i=0}^n E\{f_i(y_i)\} $. The method cannot determine ICs that are Gaussian distributed since their pdfs are symmetrically distributed and therefore cannot be distinguished.

The approach to identify the unknown ICs $s_i$ and the mixing matrix $\mathbf A$ is based on the Central Limit Theorem. This says that the pdf of a sum of independent variables tends towards a Gaussian distribution. Hence, the pdf of several independent variables is always more Gaussian than that of a single variable. Therefore, the search for one IC is performed by maximising the non-Gaussianity of an estimated component. 

Two possible measures of Gaussianity are the Kurtosis and the Negentropy. The Kurtosis is defined as the normalised fourth moment of a variable: $\rm{kurt}(y)=E\{y^4\}-3E\{y^2\}^2$,which is zero for a Gaussian distributed variable. The Negentropy is a slight modification of the Entropy: $H(y)=-\int f(y)\log(f(y))dy$. $H(y)$ reaches its maximum for a Gaussian variable since it is the most random and 'disordered' distribution. The Negentropy is defined as $J(y)=H(y_{\rm{gauss}})-H(y)$ to set the quantity to zero for a Gaussian variable and make it non-negative. However, the Negentropy is computationally hard to determine therefore an approximation of the Non-Gaussianity, which uses the Kurtosis, is applied.
\subsection{Application}
\begin{figure}
\includegraphics[width=.5\textwidth]{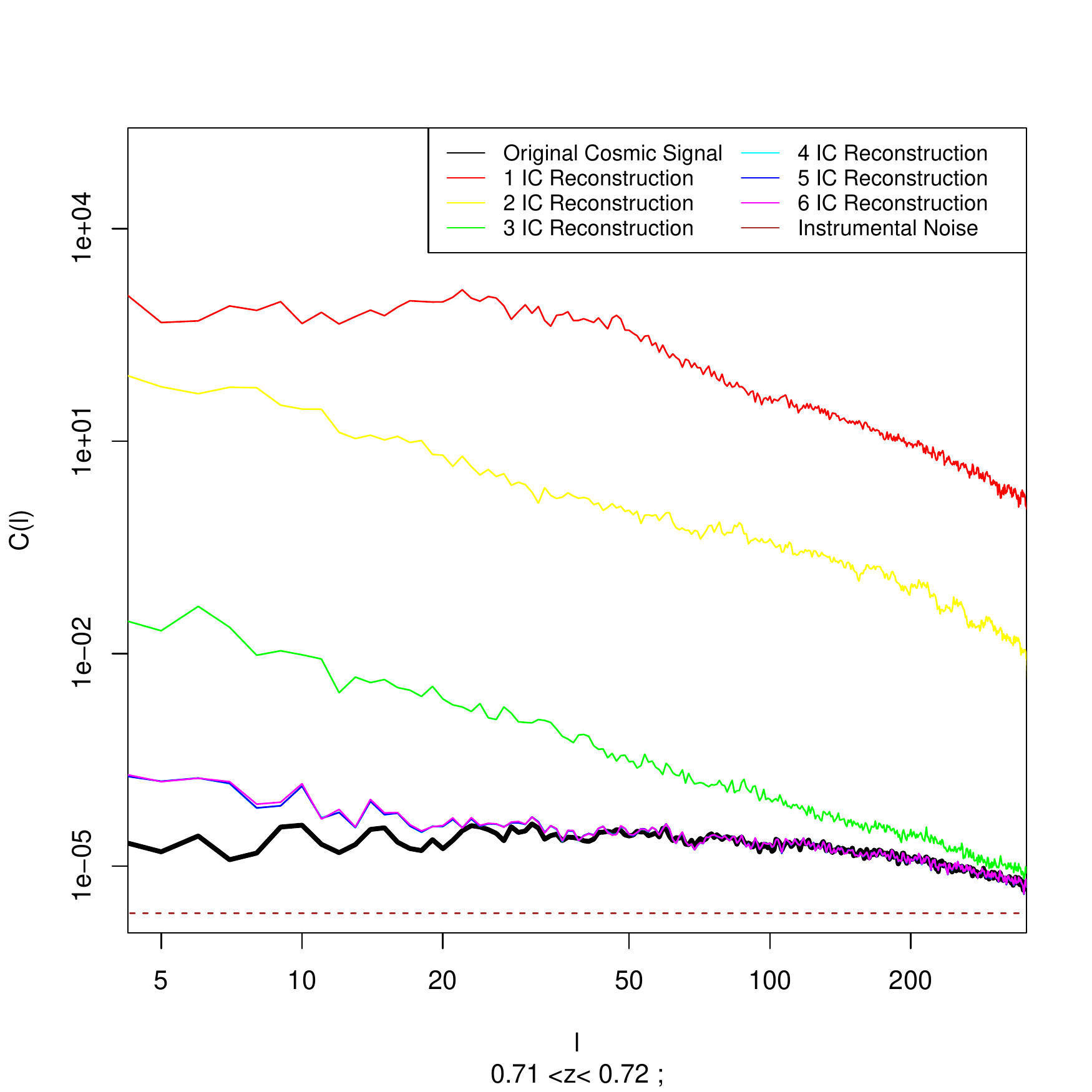}
\caption[Power spectra of reconstructed signal for different numbers of ICs]{Power spectra of the original cosmological signal (black), the receiver noise (brown), the recovered cosmological signal after the foreground removal (coloured) for number of independent components (IC) 1 (red), 2 (yellow), 3 (green), 4 (light blue), 5 (dark blue) and 6 (pink).}
\label{fig:fastica_diffICA}
\end{figure}
In our application of the \textsc{fastica} method, one measurement $x_i$ is a Healpix map at a given frequency slice with $N_{\rm{pix}}$ entries of the HI intensity map. Therefore the whole input vector is a matrix of dimension (Number of frequencies) $\times$ (Number of pixels per map). In our analysis, we binned the simulation in $N_{\rm{maps}}=160$ frequency shells and used the Healpix resolution $N_{\rm{pix}}=12\cdot512^2$ such that we can measure the angular power spectum to $l_{\rm{max}}\simeq 500$. The ICs are hence Healpix maps of the same resolution. As explained in Section \ref{1sec:noise} the receiver noise is a realisation of a Gaussian probability density function, therefore the \textsc{fastica} does not consider it to be part of the ICs. The Galactic foregrounds can be up to five magnitudes higher than the original cosmological signal. Hence, the method only identifies the Galactic foregrounds as the ICs and the difference between reconstruction and input map is the recovered signal plus the receiver noise. However, the mean temperature $T_{\rm{mean}}$ of the cosmological signal simulations is a smooth function of redshift and is therefore incorporated into the ICs of the analysis. The residual maps require to be renormalised to the $T_{\rm{mean}}$ of the input maps to ensure consistency.

We performed the foreground removal with $N_{\rm{IC}}\in\{1,2,3,4,5,6\}$ as can be seen in Fig.~ \ref{fig:fastica_diffICA}. 
It is evident that an analysis with less than 4 independent components does not remove the foregrounds sufficiently, such that they leak into the reconstructed cosmological signal. For the case of 4 IC or more, we see that the cosmological data is well recovered for multipoles larger than 50. The Galactic foreground is particularly high in the Galactic plane and therefore contaminates the large-scale structure reconstruction. We chose 4 independent components to be optimal in our analysis. In the following, we test the effect of different masking in order to reduce the large-scale contamination.
\subsection{Masking}
\label{3sec:mask}
\begin{figure*}
\centering
\subfigure[Constant Sky Cut]{         
                  \centering  
\includegraphics[width=0.5\textwidth, clip=true, trim= 0 0 0 0]{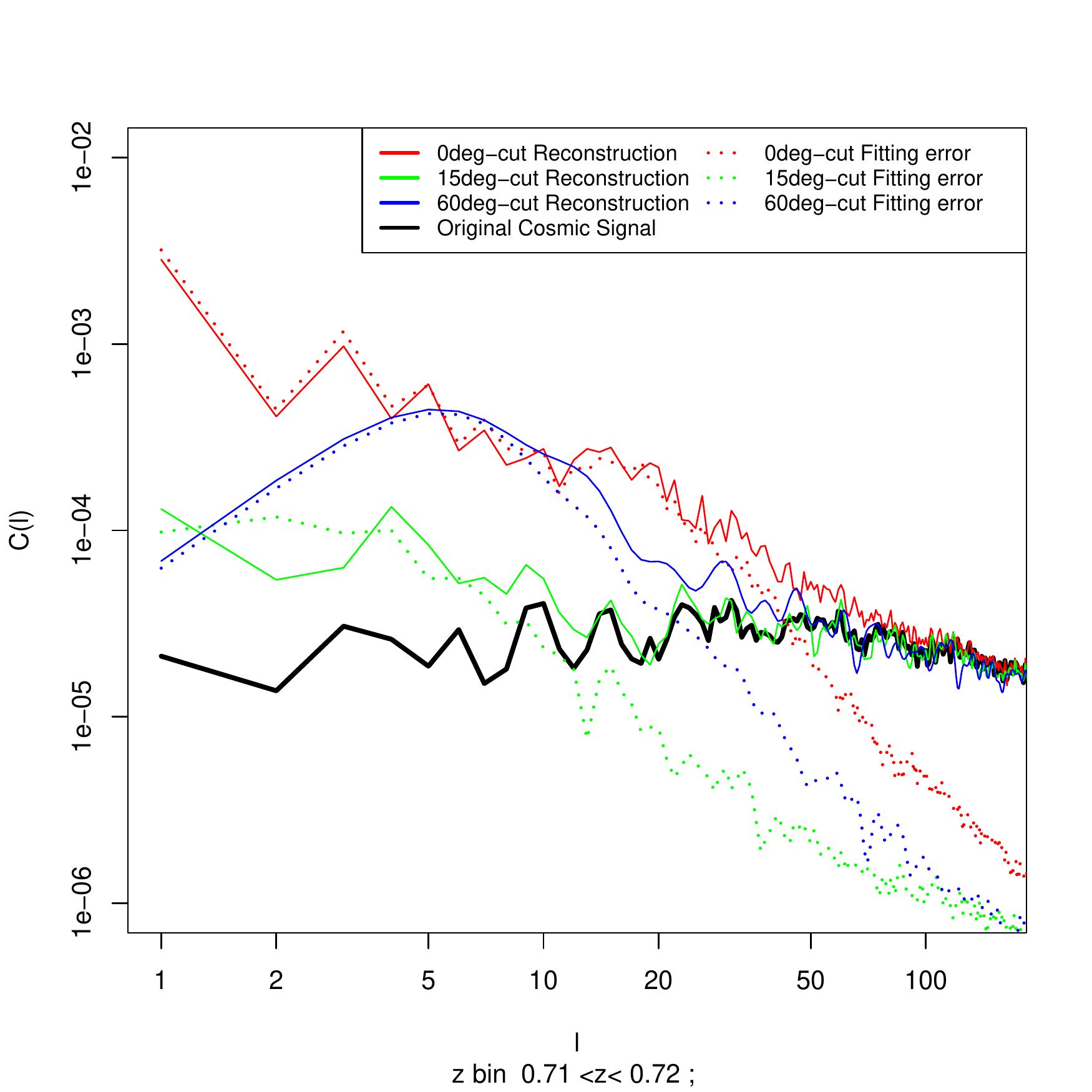}
\label{fig:CS_Fiterr_const}
}%
\centering
\subfigure[Temperature Cut]{         
                  \centering  
\includegraphics[width=0.5\textwidth, clip=true, trim= 0 0 0 0]{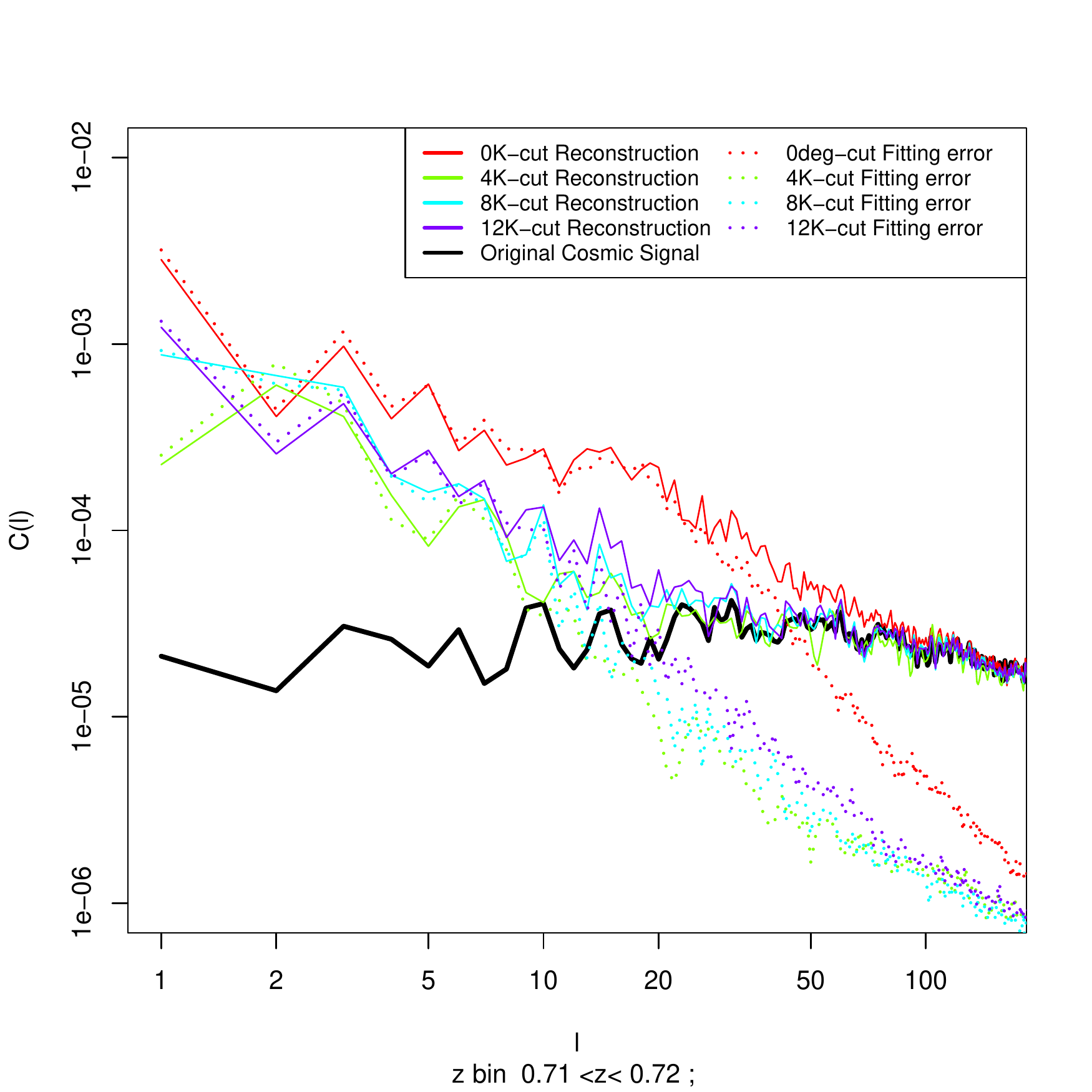}
\label{fig:CS_Fiterr_temp}
}
\caption[Power spectra of reconstructed cosmic signal and fitting error for different masks]{Power spectra of the original cosmological signal (black), the reconstructed cosmic signal (solid coloured) and the fitting error (dotted coloured) for different masks with no Galactic cut (red) and Galactic cuts of $\theta=15\rm{deg}$ (green) and $\theta=60\rm{deg}$ (blue) with 4 independent components in the left panel. The same original power spectrum (black) is shown in the right panel for different masks with threshold on the foreground temperature $T_{\rm{max}}=0K$ (red), $T_{\rm{max}}=4K$ (green), $T_{\rm{max}}=8K$ (light blue) and $T_{\rm{max}}=12K$ (dark blue) with 4 independent components. This figure shows results for the $0.71<z<0.72$ frequency slice.}
\label{fig:CS_Fiterr_mask}
\end{figure*}
\begin{figure}
\includegraphics[width=0.5\textwidth, clip=true, trim=  0 0 0 0]{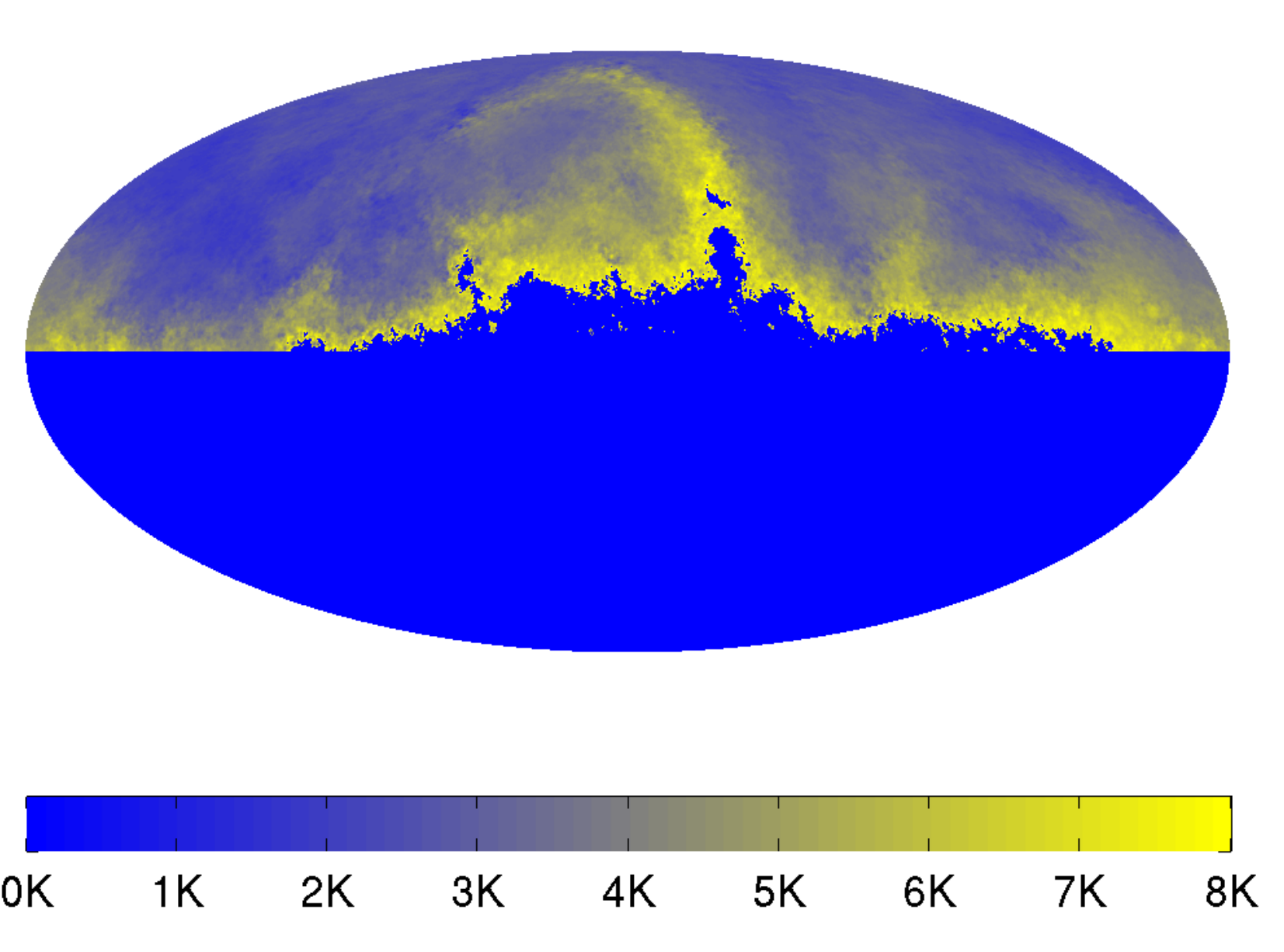}
\includegraphics[width=0.5\textwidth, clip=true, trim=  0 0 0 0]{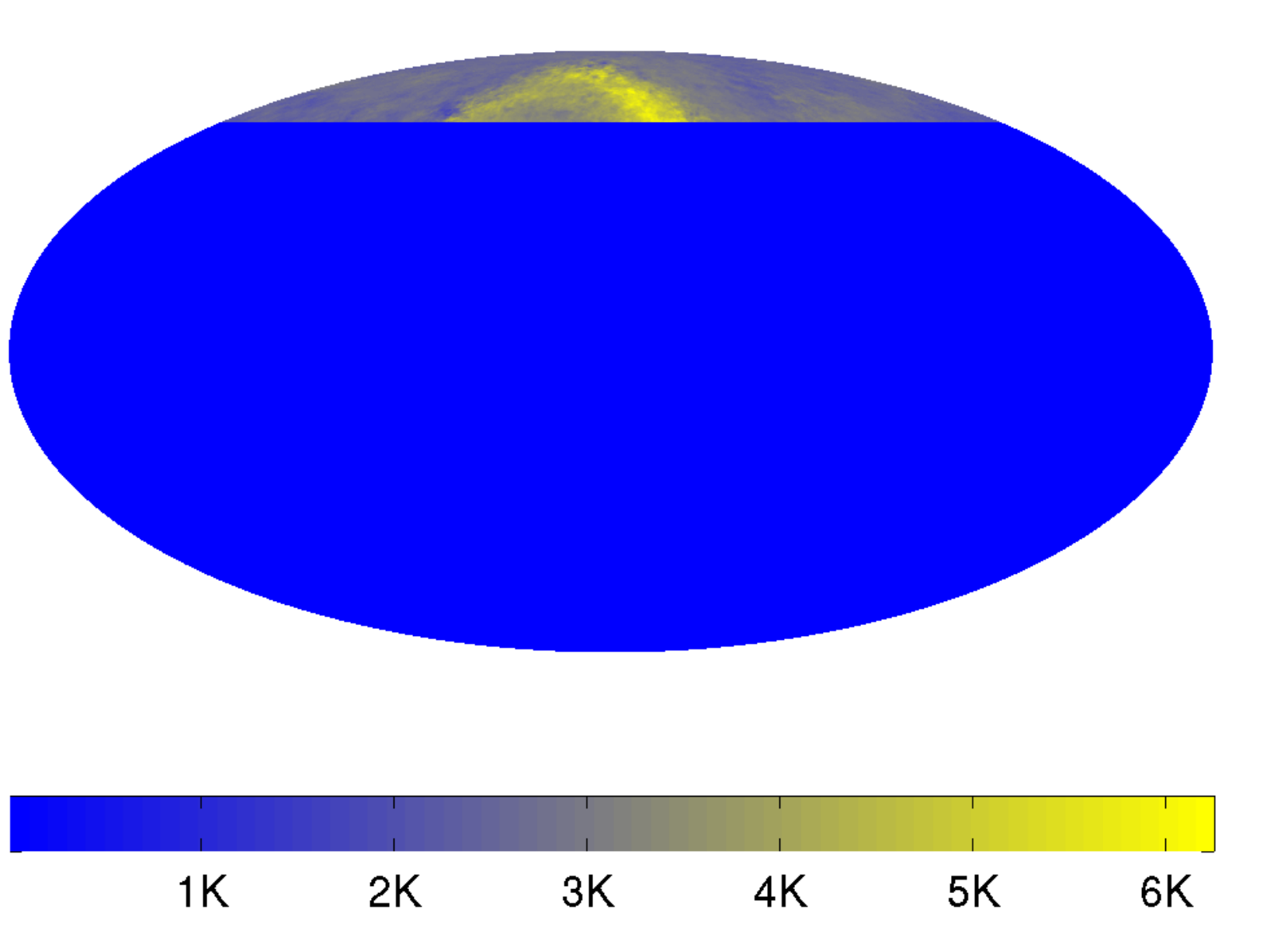}
\caption[Masked Galactic foreground]{Masked Galactic foreground where pixels with $T>8K$ are removed in the upper map and with a constant Galactic latitude sky cut of $60^\circ$ in the lower map, shown for the frequency slice $\nu=829\rm{MHz}$}
\label{fig:mapsfastica}
\end{figure}
The radio emission from the Milky Way is particular high in the Galactic plane in comparison to the emission at higher latitudes. This causes the power spectrum reconstruction to be poorer on large-scales. It is convenient to mask the input data to decrease the contamination. The aim is to strike a balance between including the largest possible survey area and removing the areas of the most dominating foreground emission. We examined the output of \textsc{fastica} with unmasked input data, as seen in the previous section. We calculated the fitting error of the temperature, which is the difference between the reconstructed cosmological signal $T_{\rm{rec}}$ and the input cosmological signal $T_{\rm{CS}}$ plus noise $T_{\rm{NO}}$: $\Delta T_{\rm{fit}}=T_{\rm{rec}}-T_{\rm{CS}}-T_{\rm{NO}}$. We considered the power spectrum of the reconstructed cosmological signal in comparison with the power spectrum of the fitting error, and aimed to reduce the latter. For our mask testing, we chose a medium redshift bin of $z=0.715$.

In creating an optimal mask, we considered two approaches. First, we created a mask with a constant cut in Galactic latitude and applied it to the input data. We examined the cases with a $15^\circ$ and a $60^\circ$-cut. The first case aims to remove the strongest foregrounds in the Galactic plane. The second case, the $60^\circ$-cut, reduces the observed sky area to about 2760 square degrees. This simulates a very conservative survey in the North Galactic Pole which is feasible with an SKA pathfinder.
We performed the \textsc{fastica} on the two cut-sky simulations and find that the masking enhances the quality of reconstruction on larger scales in comparison with an unmasked analysis, as shown in Fig.~ \ref{fig:CS_Fiterr_const}. The fitting error decreases in the analysis. However, the constant limit in latitude removes large areas where the foreground temperature is relatively low and hence information on the large-scales of the matter distribution is lost. 

We circumvented the aforementioned loss of information by creating a mask by applying a threshold $T_{\rm{max}}$ on the temperature of the Galactic foreground. We mask the pixels above $T_{\rm{max}}$ and thereby do not lose 'good' pixels with low latitude. In Fig.~ \ref{fig:CS_Fiterr_temp} the outcome of applying different foreground thresholds to the data is pictured. The error for the masked \textsc{fastica} results is always smaller than for an unmasked experiment. We conclude that the optimal threshold is $T_{\rm{max}}=8K$, where the fitting error is relatively small while preserving a large sky area. The fitting error is less than the power spectrum signal for $\ell >20$, which is a great improvement compared to the case of no masking where systematic errors were dominant for $\ell <50$. 

For the cosmological analysis in Sec. \ref{sec-chi2} we consider two cases; the constant latitude cut of $60^\circ$ to mock an SKA precursor experiment and the half-sky simulation masking in foreground temperature above $8K$. The masked Galactic foregrounds for the different cases are shown in Fig.~ \ref{fig:mapsfastica}.

\subsection{Residual Projection}
\label{3sec:residual}
To evaluate the performance of the \textsc{fastica}, we calculate the amount of Galactic foreground and recovered signal that leak into each other. Each element (foreground, cosmic signal, noise ) constituting the input data can be projected onto the reconstructed elements via the mixing matrix. We can therefore split the fitting error into the contributions of the single constituents and deepen our understanding of the sources of the systematic errors.

In the case of the Galactic foreground this is done via the following equation. The residual that leaks into the recovered signal and noise is:
\begin{equation}
R_{\rm{fg}}=\rm{fg}-(\mathbf A( \mathbf A^T \mathbf A)^{-1}\mathbf A^T) \rm{fg}.
\label{eq:fgres}
\end{equation}
In this equation, $\rm{fg}$ is the input Galactic foreground data from which we subtract the foreground maps projected onto the mixing matrix $A$. Similar to Eq.~ \ref{eq:fgres}, we define the noise (no) leakage and the noise plus signal (nocs) leakage as:
\begin{equation}
R_{\rm{no}/\rm{nocs}}=(\mathbf A( \mathbf A^T \mathbf A)^{-1}\mathbf A^T) (\rm{no}/\rm{nocs}).
\label{eq:leakage}
\end{equation}
We quantify the leakage maps of the foreground and noise and cosmic signal by calculating their power spectra and comparing these to the original power spectrum of the input simulation. If the foreground leakage is significantly lower than the input power spectra of noise and cosmic signal, the foreground removal worked successfully. Another important goal is that the noise and cosmic signal leakage is lower than the original power spectrum, otherwise too much of the signal gets lost into the foreground reconstruction.

\begin{figure}
\includegraphics[width=0.5\textwidth, clip=true, trim=  0 0 0 0]{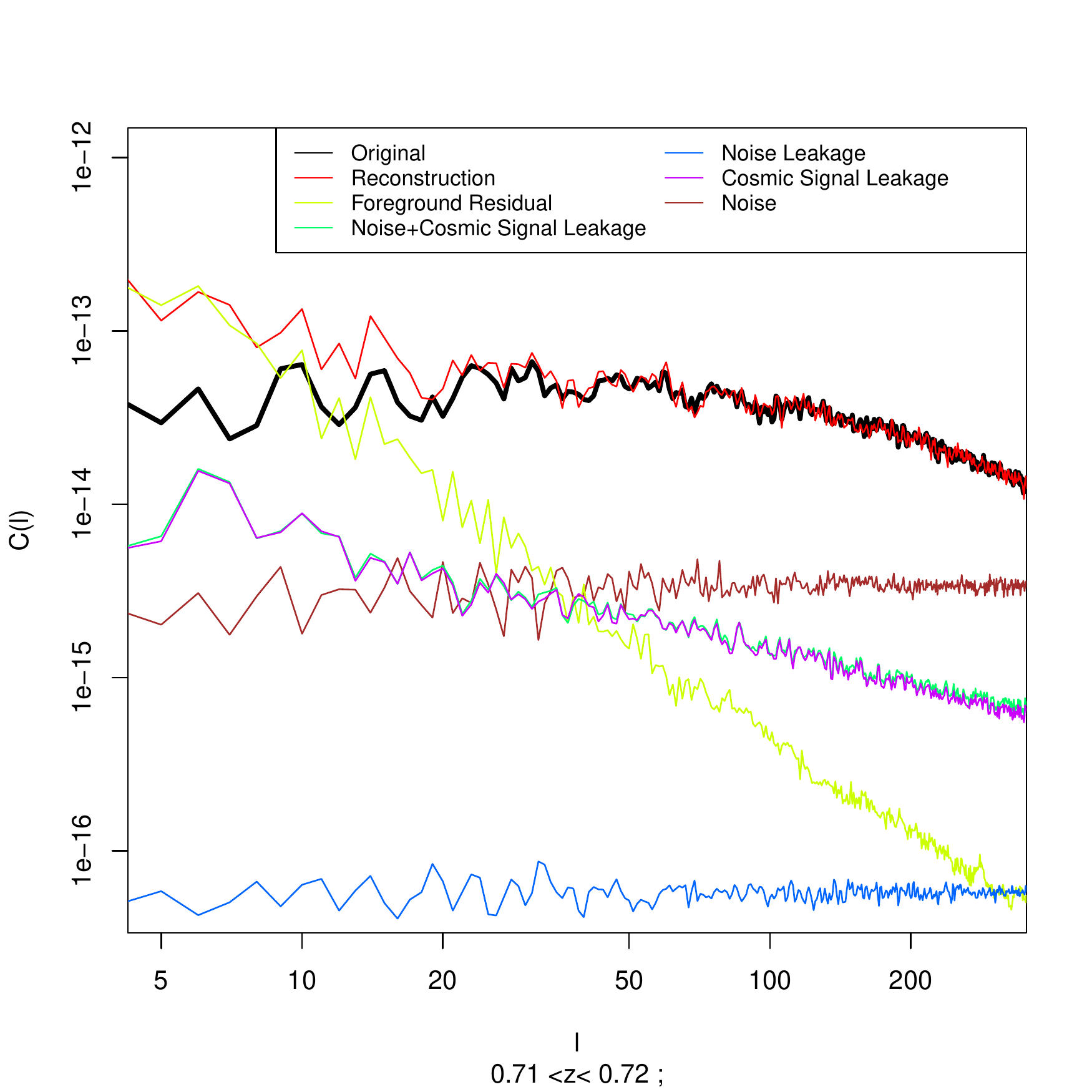}
\caption[Power spectra of foreground and cosmological signal+noise leakage for different masks]{Power spectra of the original cosmological signal (black), the receiver noise (brown), the reconstructed cosmological signal(red), the residuals of the Galactic foreground that leak into the recovered signal (yellow), the noise leakage into the foreground reconstruction (blue), the cosmological signal leakage into the foreground reconstruction (purple), the cosmological signal plus noise leakage into the foreground reconstruction (green) with 8K-cut-off mask and 4 independent components. Results are for the frequency slice $0.71<z<0.72$.}
\label{fig:fgres_nocs_mask}
\end{figure}

In Figure \ref{fig:fgres_nocs_mask}, we show the residual leakage for the mask with cut in temperature $T_{\rm{max}}=8K$ and 4 independent components. The variance of the noise and cosmic signal leakage is lower than the original power spectrum on all scales and we can be confident that we recover most of the information of the original signal. The contaminations of the foreground are only significant on small multipoles and become very small for large multipoles. This confirms the previous findings that the large-scales are contaminated by residuals of the Galactic foreground.


%

\section{Cosmological Analysis}
\label{sec-chi2}
We now consider the cosmological implications of the systematic errors in the power spectrum induced by the foreground subtraction by fitting model power spectra and baryon oscillations in each slice. In these analyses the power spectrum is estimated from rebinned temperature maps with redshift width $\delta z\approx 0.05$. We use $N_{\rm{zbin}}=27$ redshift shells within the range $0.01<z<1.49$.

\subsection{Relative Systematic Errors}
\label{sec-sys}

\begin{figure*}
\centering
\subfigure[8K-temperature sky cut]{         
        \centering  
	\includegraphics[width=0.5\textwidth]{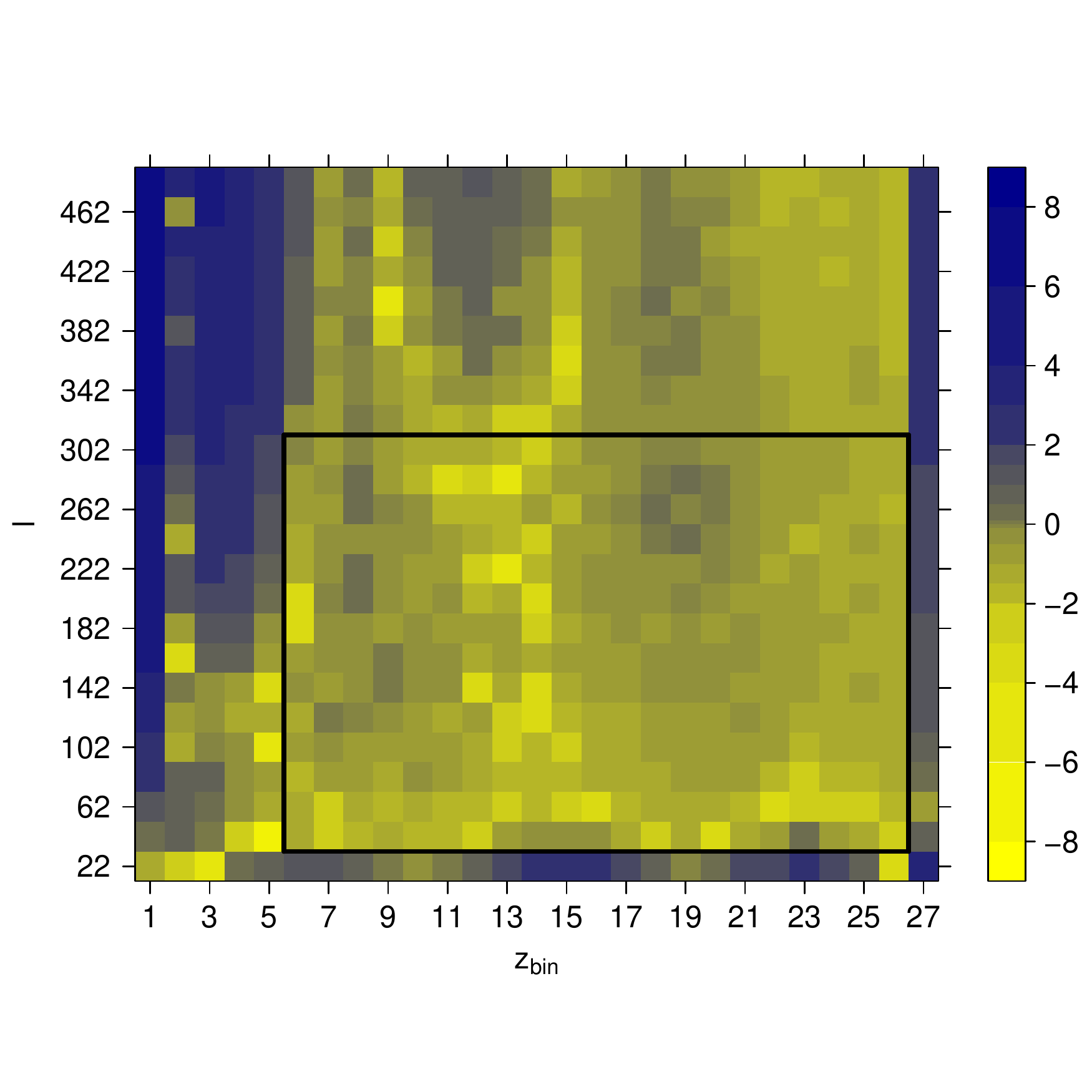} 
	\label{fig:clsysoverstat_8K}
	}%
	\subfigure[60 degree sky cut]{         
        \centering  
	\includegraphics[width=0.5\textwidth]{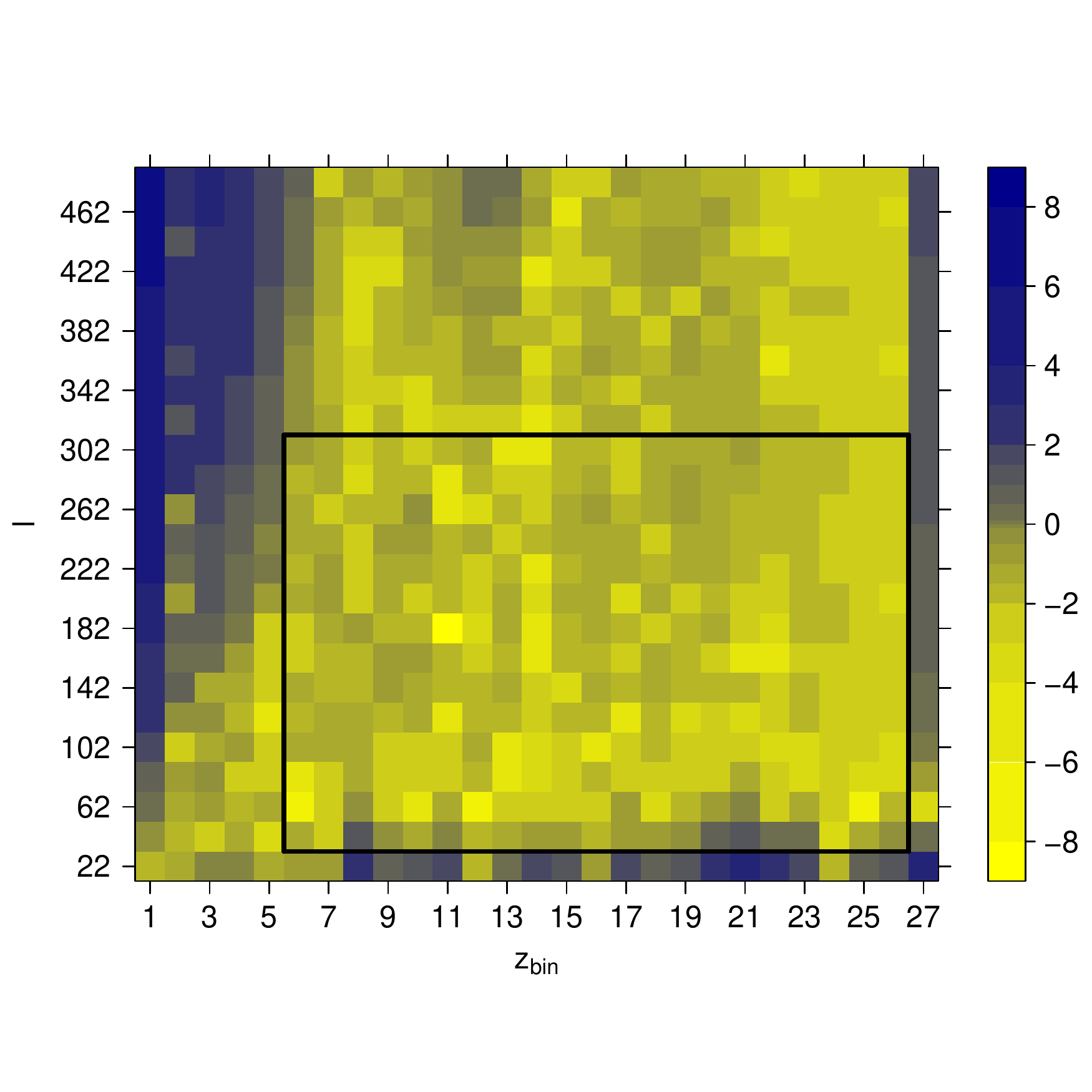} 
	\label{fig:clsysoverstat_60deg}
	}
\caption{Systematic errors in the power spectrum caused by the foreground subtraction with 4 ICs divided by the statistical errors described in Eq.~ \ref{noiseeq} as a function of redshift on the x-axis and multipole on the y-axis shown in a logarithmic scale. The right panel shows the analysis of the 8K-temperature cut sky maps and the left panel the constant cut in latitude with 60 degrees.}
\label{fig:clsysoverstat}
\end{figure*}
In Sec.~\ref{sec-fastica}, we have shown that the systematic errors compared to the absolute value of the power spectra are relatively small. In the following, we consider the systematics relative to the statistical errors in the power spectrum to understand how relevant they are in the cosmological analysis.

In Fig.~ \ref{fig:clsysoverstat}, the systematic errors in the power spectrum defined as $| C_{\rm{Sys}}(\ell ) |=| C_{\rm{orig}}(\ell )-C_{\recdata}(\ell )|$ are divided by the statistical errors given by Eq.~\ref{noiseeq}. The relative systematics are given as a function of redshift on the x-axis and multipole on the y-axis in a natural logarithmic scale. The errors are binned with $\delta \ell =20$. The large-scale contamination can be seen as the high values in the bottom row of the matrix plots. The redshift bins on the edges of the data cube can also not be well recovered by the \textsc{fastica}. In addition, the removal leaves contaminations on the small scales of the power spectrum, particularly for small redshifts. The systematic errors on small scales are not very high in power, however, relative to the Cosmic Variance, which is very small for high $\ell$, they become significant. 
The mean relative systematic errors in Fig.~\ref{fig:clsysoverstat_8K} is about 20 sigma, which is caused by the high errors at the edges of the data cube. The systematics in the black box, however, lie between 0 and 1.3$\sigma$ with a mean error of 0.5$\sigma$.
The behaviour of the relative systematics as a function of $z$ and $\ell $ is very similar for the two different cases of masking. For the $60^\circ$ sky cut in Fig.~ \ref{fig:clsysoverstat_60deg}, the relative systematics are of smaller power because the statistical error scales with sky area, which is a factor of $6.7$ smaller than for the $8K$- temperature cut shown in Fig.~ \ref{fig:clsysoverstat_8K}.

In the following cosmological parameter estimation, we consider the relatively uncontaminated power spectrum in the range $20<l<300$ in the redshift bins 5 to 26, outlined by the black square in Fig.~ \ref{fig:clsysoverstat}. 
\subsection{Cosmological Parameter Estimation}
\begin{figure}
\centering
\subfigure[8K-temperature sky cut]{         
                  \centering 
\includegraphics[width=0.5\textwidth, clip=true, trim=  0 20 10 10]{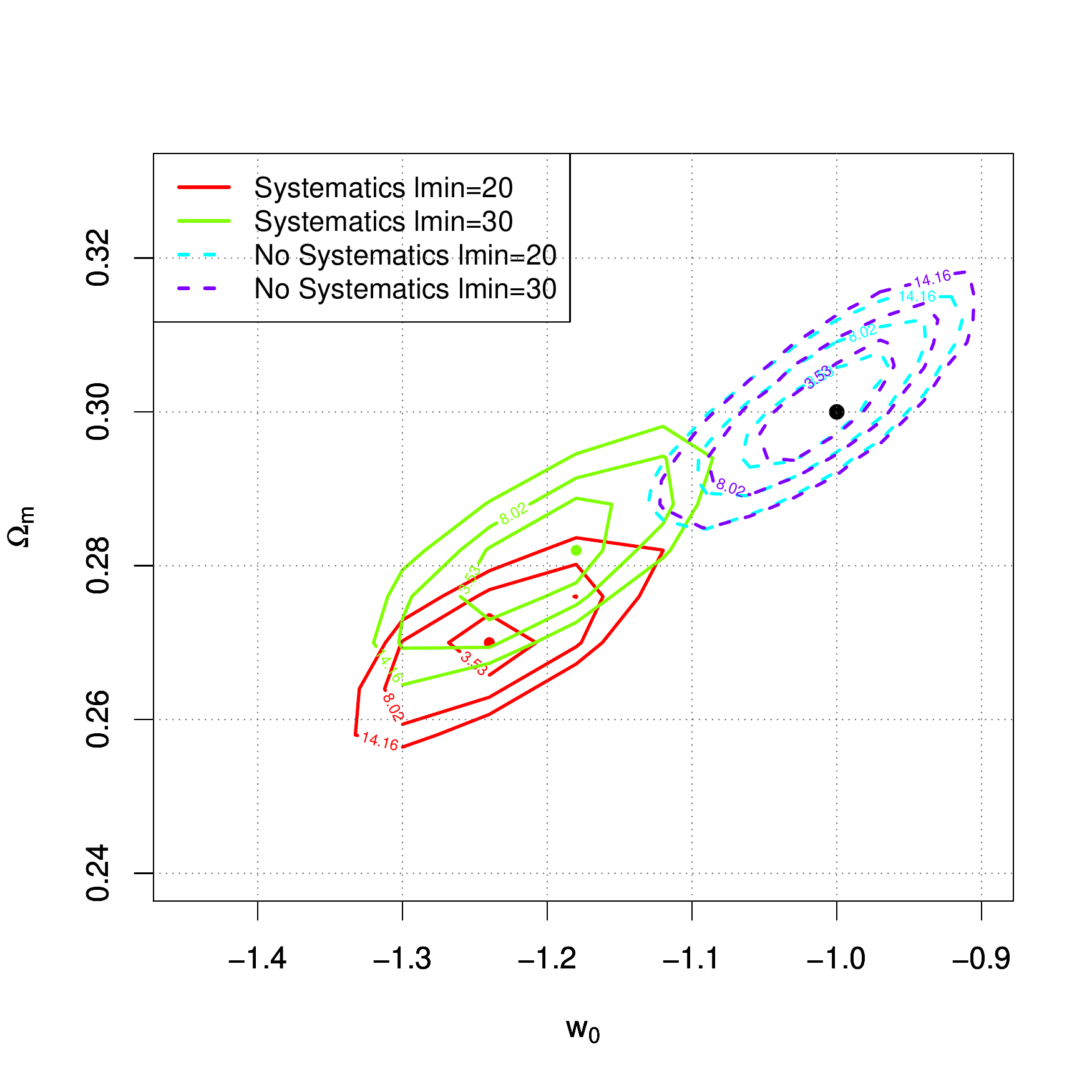}
\label{fig:chi2_8K}
}%

\subfigure[60 degree sky cut]{         
                  \centering 
\includegraphics[width=0.5\textwidth, clip=true, trim=  0 20 10 10]{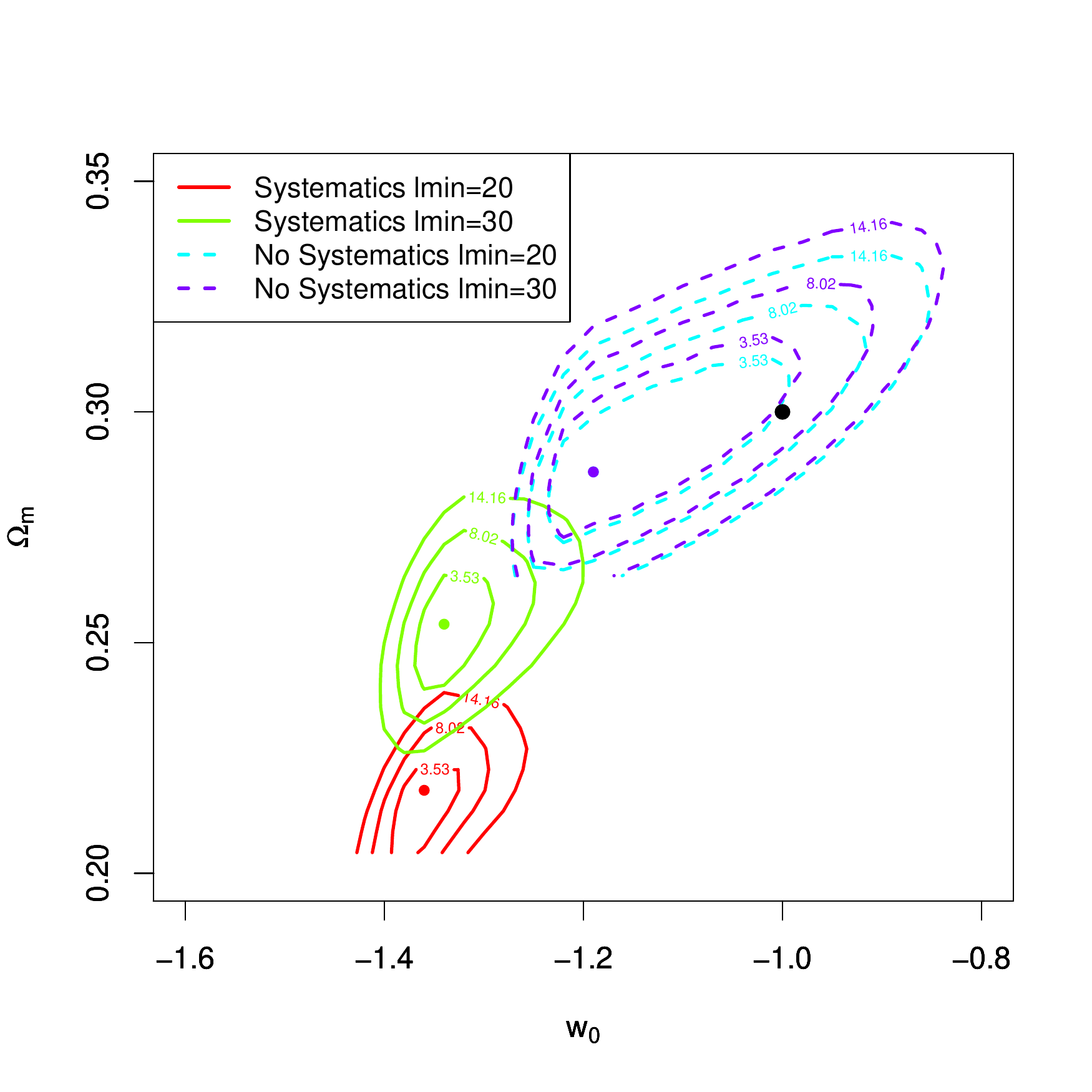}
\label{fig:chi2_60deg}
}
\caption{1, 2 and 3-sigma contours of the $\chi^2$ values of $\Omega_m$ and $w_0$ marginalized over galaxy bias, for the uncontaminated power spectrum (dashed lines) and the power spectrum including the systematics of the foreground removal (solid lines) for analysis run with different $l_{\rm{min}}$. The black dot shows the fiducial cosmological model. The coloured dots mark the best-fit values found by the parameter estimation for the different cases. The foreground subtraction biases the recovery of the cosmological parameters.}
\label{fig:chi2}
\end{figure}
First we estimate the best-fitting cosmological parameters $\bmath p$ of the power spectrum. We calculate the $\chi^2(\bmath p)$ value of the parameters on a grid in parameter space. In the following, the 1,2 and 3-sigma contours for $\bmath p=(\Omega_m,w_0)$ with different choices of multipole range are presented. We marginalise over the bias parameter $b$. The remaining cosmological parameters are unchanged and set to the fiducial model described in Sec. \ref{2sec:theory}. As motivated in Sec.~\ref{sec:cov}, we use a diagonal covariance matrix, which results in the following simple form of the $\chi^2(\bmath p) $
\begin{align}
\chi^2(\bmath p)= & \sum_{i=1}^{N_{\rm{zbin}}}\sum_{\ell '=\ell _{\rm min}}^{\ell _{\rm max}} \frac{ (C_{\rm{ data}, z_i}(\ell )-C_{\rm{theory}, z_i}(\ell , \bmath p))^2} { \rm{Cov}_{z_i}(\ell ,\ell ) }
\end{align}

We want to isolate systematics due to the foreground subtraction. For that reason, the data power spectrum used in the cosmological parameter estimation is based on the fiducial theoretical power spectrum, where we add cosmic variance calculated with the lognormal realisations and the systematics of the foreground removal as given in Section \ref{sec-sys}. This reads as
\begin{equation}
C_{\rm{ data}, z_i}(\ell )=C_{\rm{theory}, z_i}(\ell )+C_{\rm{var},z_i}(\ell )+C_{\rm{sys},z_i}(\ell )
\end{equation}
where the cosmic variance is computed via $C_{\rm{var},z_i}(\ell )=C_{\rm{log},z_i}(\ell )-\bar C_{\rm{log},z_i}(\ell )$ and the systematics as described in Sec. \ref{sec-sys} as $C_{\rm{sys},z_i}(\ell )=C_{\rm{\recdata},z_i}(\ell )-C_{\rm{orig}, z_i}(\ell )$.

The choice of $\ell_{\rm{min}}$ is an important consideration for the analysis, since the large-scales of the cleaned maps show relics of the Galactic foreground as explained in the previous Section. If we choose a $\ell_{\rm{min}}<20$, no useful parameter estimates can be computed since the systematic error on large-scales are too dominant. For that reason, in Fig.~ \ref{fig:chi2} we present the parameter contours for $\ell_{\rm{min}}=\{20,30\} $.  The maximum multipole is held constant as $\ell=300$. As a reference, the parameter contours for an analysis setting $C_{\rm{sys}}(\ell) =0$ are shown with dashed lines where the contours recover the input cosmology marked with a black dot.

In Fig.~ \ref{fig:chi2_8K}, the multipole binning is chosen as $\delta \ell=2$ due to half-sky coverage of the simulated intensity maps. The green and red solid lines mark the 1,2 and 3-sigma parameter constraints which are markedly biased by the foreground removal systematics compared to the dashed lines of the original reference data. However, we note that increasing $l_{\rm{min}}$ to 30 reduces the parameter bias.

For the $60\deg\rm{-cut}$ masking of the intensity maps, the multipoles are more highly correlated due to the small sky cut, hence we use multipole binning $\delta \ell=10$. This leads to weaker constraints due to the larger Cosmic Variance error. Apart from higher inaccuracies in $\Omega_{\rm{m}}$ due to the sky cut, the systematic errors induce a similar bias on the contours as in Fig.~~\ref{fig:chi2_8K}. We conclude that the systematics of the foreground removal prevent  the recovery of unbiased cosmological parameters.
\subsection{Baryon Acoustic Oscillation Fit}
\begin{figure}
\begin{center}
\resizebox{0.5\textwidth}{!}{\rotatebox{270}{\includegraphics{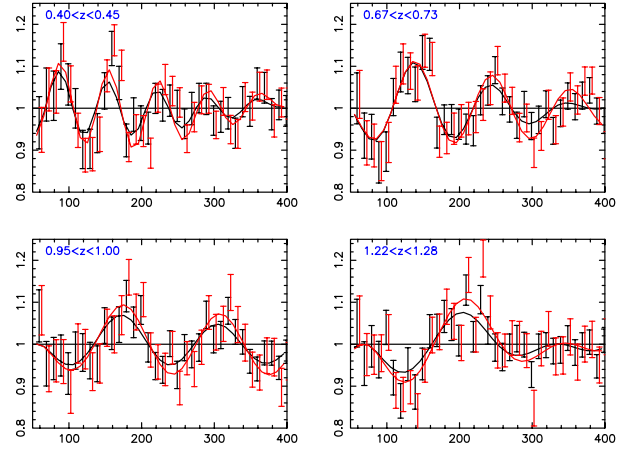}}}
\end{center}
\caption{The BAO wiggles shown in example redshift slices. Black and red points show measurements using data without any foregrounds, and following the addition and removal of foregrounds, respectively. The underlying lines give the fitted model to the data points.}
\label{figbao1}
\end{figure}
\begin{figure}
\begin{center}
\resizebox{0.5\textwidth}{!}{\rotatebox{270}{\includegraphics{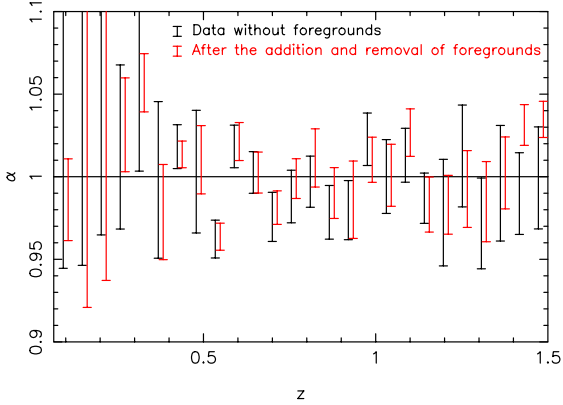}}}
\end{center}
\caption{The BAO distortion scale $\alpha$ measured in a series of redshift slices, such that $\alpha = 1$ equates to the input cosmological model.  Black and red points show measurements using data without any foregrounds, and following the addition and removal of foregrounds, respectively.}
\label{figbao2}
\end{figure}
We investigate whether the systematic distortions imprinted in the
power spectrum shape by foreground removal biased the recovery of
cosmic distances by measuring BAO in
each redshift slice of the dataset using the half-sky masked intensity mapping data.
We fitted a BAO model to the
angular power spectra both including and excluding the systematic
contribution $ C_{{\rm sys}}(\ell)$ that results from the
addition and removal of foregrounds.  Our adopted BAO model followed
equation 7 in Seo et al.\ (2012):
\begin{equation}
C_{{\rm obs},z_i}(\ell) = B_{z_i}(\ell) C_{{\rm theory},z_i}(\ell/\alpha)
+ A_{z_i}(\ell) .
\end{equation}
In this equation, $\alpha = D_A(z_i)/D_{A,{\rm fid}}(z_i)$ is the
fitted scale distortion parameter, applied to the fiducial model
angular power spectrum in the $i$th redshift slice $C_{{\rm
theory},z_i}(\ell)$, which describes the best-fitting angular diameter
distance, $D_A(z_i)$, relative to its value in the fiducial cosmology,
$D_{A,{\rm fid}}(z_i)$.  We used the input cosmology of the simulation
as the fiducial cosmology, such that $\alpha = 1$ implies that the BAO
scale is recovered without bias.  We constructed the fiducial angular
power spectrum in each redshift slice, $C_{{\rm theory},z_i}(\ell)$, by
projecting the input spatial power spectrum of the simulation over the
redshift slice.  The terms $A_{z_i}(\ell) = \sum_{j=0}^{N_A} A_j \,
\ell^j$ and $B_{z_i}(\ell) = \sum_{j=0}^{N_B} B_j \, \ell^j$ are
polynomials whose coefficients we marginalize over to describe general
power-spectrum shape distortions.  We find that a 2nd-order
polynomial, $N_A = N_B = 2$, produces a good description of the
systematic contribution.

In each redshift slice we binned the angular power spectrum
measurements in multipole bands $\delta \ell = 10$.  The covariance
matrix of the measurements, assumed diagonal given the wide sky area
covered ($f_{\rm sky} = 0.5$) and the tests described above, was
constructed using the description in Sec.~\ref{sec:cov}.  The
covariance matrix of the foreground-subtracted measurements was
obtained by adding and removing foregrounds from each realization.  We
performed the fits by varying the BAO parameter $\alpha$ over a grid
for each redshift slice, and for each value of $\alpha$ using a
downhill simplex technique to minimize the $\chi^2$ statistic varying
the polynomial coefficients $A_j$ and $B_j$.  We fitted the model over
the multipole range $50 < \ell < 400$.

The BAO wiggles of an unbiased power spectrum and of the one after foreground removal are plotted in Fig.~\ref{figbao1}, where it can be seen that the position of the wiggles remain unchanged. 
Fig.~\ref{figbao2} compares the measurements of $\alpha$ from data
without any foregrounds, and following the addition and removal of
foregrounds.  We note that our capacity to extract the BAO signature
is not biased by the foreground removal technique.
\section{Conclusions}
\label{sec-con}
We have created the most realistic simulation to date of a future half-sky HI intensity mapping experiment, that includes HI in galaxies with an appropriate mass function and clustering, Galactic foregrounds with appropriate frequency correlations, and receiver noise for a range of different experiments.
We apply foreground removal to this map using \textsc{fastica}, and reconstruct the cosmological signal.  We show that 4 independent components is the optimal reconstruction choice, and consider applying masks based on either Galactic latitude or foreground intensity level.  Best results are obtained after masking pixels with brightness temperatures $>8\rm K$.
With this method, the reconstructed cosmological power spectra contain scale-dependent systematic errors compared to the input cosmological model which are on average 50\% of the statistical error for the multipole range $20<\ell<300$ in the redshift bins considered in the analysis.
The systematics bias the recovery of cosmological parameters that depend on the overall shape of the power spectrum. These systematic errors are not ameliorated by imposing a stricter Galactic plane or foreground intensity cut. The validity of the presented conclusions is restricted to the analysis with the chosen foreground subtraction method. The systematic errors might reduce and exhibit different behaviour on small and large-scales for other subtraction methods.

Furthermore, the addition and removal of foregrounds imprints significant and complex correlations between power spectrum multipoles that are not present in the raw cosmological signal.  In order to understand such correlations, future intensity-mapping experiments will require a large suite of realistic simulations.

Despite the presence of these systematic errors we show that the signature of BAOs may nonetheless be recovered in an unbiased manner from the intensity mapping survey, using standard techniques that combine a template acoustic-oscillation model with broad-band polynomial shape terms, with free coefficients which are marginalized.

We conclude that, for a range of intensity mapping experiments, \textsc{fastica} cannot extract the power spectrum of the cosmological signal in an unbiased manner.  However, the BAO signature can nonetheless be obtained in an unbiased way, and we therefore suggest that such surveys will be extremely powerful for mapping cosmic distances in a redshift range which is difficult for optical surveys to access.

In future work, various foreground removal techniques and their imprints into the recovered cosmological signal should be compared. Furthermore, polarized data and particularly the issue of polarization leakage requires investigation in simulated datasets. In addition, we are planning to apply \textsc{fastica} to existing intensity maps of the GBT \citep{2013MNRAS.434L..46S}.

\section*{Acknowledgments}
We appreciate the help of Ian Heywood with the SKA simulations. We would like thank Matt Jarvis, Reza Ansari, Jason McEwen and Boris Leistedt for very useful discussions and thoughtful comments.
LW is supported by the IMPACT fund. FBA acknowledges the support of the Royal Society via a University Research Fellowship. CB acknowledges the support of the Australian Research Council through the award of a Future Fellowship. We are also grateful for support from the Centre for All-Sky Astrophysics, an Australian Research Council Centre of Excellence funded by grant CE11000102.
 
\bibliographystyle{mn2e}
\bibliography{PhD2}

\begin{thebibliography}{}

\bibitem[\protect\citeauthoryear{A.~R.~Thompson J.
  M.~Moran}{A.~R.~Thompson}{2004}]{thompson04}
A.~R.~Thompson J. M.~Moran G. W. J.~S.,  2004, Interferometry and Synthesis in
  Radio astronomy.
WILEY-VCH Verlag GmbH \& Co. KGaA, Weinheim

\bibitem[\protect\citeauthoryear{{Abdalla}, {Blake} \& {Rawlings}}{{Abdalla}
  et~al.}{2010}]{Abdalla:2009wr}
{Abdalla} F.~B.,  {Blake} C.,    {Rawlings} S.,  2010, Mon.Not.Roy.astron.Soc.,
  401, 743

\bibitem[\protect\citeauthoryear{Abdalla \& Rawlings}{Abdalla \&
  Rawlings}{2005}]{Abdalla:2004ah}
Abdalla F.~B.,  Rawlings S.,  2005, Mon.Not.Roy.astron.Soc., 360, 27

\bibitem[\protect\citeauthoryear{{Ansari}, {Campagne}, {Colom}, {Le Goff},
  {Magneville}, {Martin}, {Moniez}, {Rich} \& {Y{\`e}che}}{{Ansari}
  et~al.}{2012}]{Ansari:2011bv}
{Ansari} R.,  {Campagne} J.~E.,  {Colom} P.,  {Le Goff} J.~M.,  {Magneville}
  C.,  {Martin} J.~M.,  {Moniez} M.,  {Rich} J.,    {Y{\`e}che} C.,  2012,
  Astronomy\&Astrophysics, 540, A129

\bibitem[\protect\citeauthoryear{{Ansari}, {Le Goff}, {Magneville}, {Moniez},
  {Palanque-Delabrouille}, {Rich}, {Ruhlmann-Kleider} \& {Y{\`e}che}}{{Ansari}
  et~al.}{2008}]{2008arXiv0807.3614A}
{Ansari} R.,  {Le Goff} J.~.,  {Magneville} C.,  {Moniez} M.,
  {Palanque-Delabrouille} N.,  {Rich} J.,  {Ruhlmann-Kleider} V.,
  {Y{\`e}che} C.,  2008, ArXiv e-prints

\bibitem[\protect\citeauthoryear{{Battye}, {Browne}, {Dickinson}, {Heron},
  {Maffei} \& {Pourtsidou}}{{Battye} et~al.}{2013}]{Battye:2012fd}
{Battye} R.~A.,  {Browne} I.~W.~A.,  {Dickinson} C.,  {Heron} G.,  {Maffei} B.,
     {Pourtsidou} A.,  2013, Mon.Not.Roy.astron.Soc., 434, 1239

\bibitem[\protect\citeauthoryear{{Bernardi}, {de Bruyn}, {Harker}, {Brentjens},
  {Ciardi}, {Jeli{\'c}}, {Koopmans}, {Labropoulos}, {Offringa}, {Pandey},
  {Schaye}, {Thomas}, {Yatawatta} \& {Zaroubi}}{{Bernardi}
  et~al.}{2010}]{2010A&A...522A..67B}
{Bernardi} G.,  {de Bruyn} A.~G.,  {Harker} G.,  {Brentjens} M.~A.,  {Ciardi}
  B.,  {Jeli{\'c}} V.,  {Koopmans} L.~V.~E.,  {Labropoulos} P.,  {Offringa} A.,
   {Pandey} V.~N.,  {Schaye} J.,  {Thomas} R.~M.,  {Yatawatta} S.,    {Zaroubi}
  S.,  2010, Astronomy\&Astrophysics, 522, A67

\bibitem[\protect\citeauthoryear{Blake, Collister, Bridle \& Lahav}{Blake
  et~al.}{2007}]{Blake:2006kv}
Blake C.,  Collister A.,  Bridle S.,    Lahav O.,  2007,
  Mon.Not.Roy.astron.Soc., 374, 1527

\bibitem[\protect\citeauthoryear{Blake, Kazin, Beutler, Davis, Parkinson
  et~al.,}{Blake et~al.}{2011}]{Blake:2011en}
Blake C.,  Kazin E.,  Beutler F.,  Davis T.,  Parkinson D.,    et~al., 2011,
  Mon.Not.Roy.astron.Soc., 418, 1707

\bibitem[\protect\citeauthoryear{Blake, Abdalla, Bridle \& Rawlings}{Blake
  et~al.}{2004}]{Blake:2004pb}
Blake C.~A.,  Abdalla F.~B.,  Bridle S.~L.,    Rawlings S.,  2004, New
  astron.Rev., 48, 1063

\bibitem[\protect\citeauthoryear{{Bottino}, {Banday} \& {Maino}}{{Bottino}
  et~al.}{2010}]{Bottino:2009uc}
{Bottino} M.,  {Banday} A.~J.,    {Maino} D.,  2010, Mon.Not.Roy.astron.Soc.,
  402, 207

\bibitem[\protect\citeauthoryear{Carilli \& Rawlings}{Carilli \&
  Rawlings}{2004}]{Carilli:2004nx}
Carilli C.~L.,  Rawlings S.,  2004, New astron.Rev.

\bibitem[\protect\citeauthoryear{Chang, Pen, Bandura \& Peterson}{Chang
  et~al.}{2010}]{Chang:2010jp}
Chang T.-C.,  Pen U.-L.,  Bandura K.,    Peterson J.~B.,  2010, Nature, 466,
  463

\bibitem[\protect\citeauthoryear{Chang, Pen, Peterson \& McDonald}{Chang
  et~al.}{2008}]{Chang:2007xk}
Chang T.-C.,  Pen U.-L.,  Peterson J.~B.,    McDonald P.,  2008,
  Phys.Rev.Lett., 100, 091303

\bibitem[\protect\citeauthoryear{Chapman, Abdalla, Harker, Jelic, Labropoulos
  et~al.,}{Chapman et~al.}{2012}]{Chapman:2012yj}
Chapman E.,  Abdalla F.~B.,  Harker G.,  Jelic V.,  Labropoulos P.,    et~al.,
  2012, Mon.Not.Roy.astron.Soc., 423, 2518

\bibitem[\protect\citeauthoryear{Chen}{Chen}{2012}]{Chen:2012xu}
Chen X.,  2012, Int.J.Mod.Phys.Conf.Ser., 12, 256

\bibitem[\protect\citeauthoryear{Condon}{Condon}{1992}]{Condon:1992rq}
Condon J.,  1992, Ann.Rev.astron.astrophys., 30, 575

\bibitem[\protect\citeauthoryear{{de Oliveira-Costa}, {Tegmark}, {Gaensler},
  {Jonas}, {Landecker} \& {Reich}}{{de Oliveira-Costa}
  et~al.}{2008}]{deOliveiraCosta:2008pb}
{de Oliveira-Costa} A.,  {Tegmark} M.,  {Gaensler} B.~M.,  {Jonas} J.,
  {Landecker} T.~L.,    {Reich} P.,  2008, Mon.Not.Roy.astron.Soc., 388, 247

\bibitem[\protect\citeauthoryear{{Di Matteo}, {Ciardi} \& {Miniati}}{{Di
  Matteo} et~al.}{2004}]{2004MNRAS.355.1053D}
{Di Matteo} T.,  {Ciardi} B.,    {Miniati} F.,  2004, Mon.Not.Roy.astron.Soc.,
  355, 1053

\bibitem[\protect\citeauthoryear{Di~Matteo, Perna, Abel \& Rees}{Di~Matteo
  et~al.}{2002}]{DiMatteo:2001gg}
Di~Matteo T.,  Perna R.,  Abel T.,    Rees M.~J.,  2002, astrophys.J., 564, 576

\bibitem[\protect\citeauthoryear{{Dillon}, {Liu} \& {Tegmark}}{{Dillon}
  et~al.}{2013}]{2013PhRvD..87d3005D}
{Dillon} J.~S.,  {Liu} A.,    {Tegmark} M.,  2013, Phys.Rev., 87, 043005

\bibitem[\protect\citeauthoryear{{Doyle}, {Drinkwater}, {Rohde}, {Pimbblet},
  {Read} et~al.,}{{Doyle} et~al.}{2005}]{2005MNRAS.361...34D}
{Doyle} M.~T.,  {Drinkwater} M.~J.,  {Rohde} D.~J.,  {Pimbblet} K.~A.,  {Read}
  M.,    et~al., 2005, Mon.Not.Roy.astron.Soc., 361, 34

\bibitem[\protect\citeauthoryear{Efstathiou}{Efstathiou}{2004}]{Efstathiou:2003tv}
Efstathiou G.,  2004, Mon.Not.Roy.astron.Soc., 348, 885

\bibitem[\protect\citeauthoryear{{Gleser}, {Nusser} \& {Benson}}{{Gleser}
  et~al.}{2008}]{2008MNRAS.391..383G}
{Gleser} L.,  {Nusser} A.,    {Benson} A.~J.,  2008, Mon.Not.Roy.astron.Soc.,
  391, 383

\bibitem[\protect\citeauthoryear{{Gong}, {Chen}, {Silva}, {Cooray} \&
  {Santos}}{{Gong} et~al.}{2011}]{2011ApJ...740L..20G}
{Gong} Y.,  {Chen} X.,  {Silva} M.,  {Cooray} A.,    {Santos} M.~G.,  2011,
  astrophys.J.Lett., 740, L20

\bibitem[\protect\citeauthoryear{{Harker}, {Zaroubi}, {Bernardi}, {Brentjens},
  {de Bruyn}, {Ciardi}, {Jeli{\'c}}, {Koopmans}, {Labropoulos}, {Mellema},
  {Offringa}, {Pandey}, {Schaye}, {Thomas} \& {Yatawatta}}{{Harker}
  et~al.}{2009}]{2009MNRAS.397.1138H}
{Harker} G.,  {Zaroubi} S.,  {Bernardi} G.,  {Brentjens} M.~A.,  {de Bruyn}
  A.~G.,  {Ciardi} B.,  {Jeli{\'c}} V.,  {Koopmans} L.~V.~E.,  {Labropoulos}
  P.,  {Mellema} G.,  {Offringa} A.,  {Pandey} V.~N.,  {Schaye} J.,  {Thomas}
  R.~M.,    {Yatawatta} S.,  2009, Mon.Not.Roy.astron.Soc., 397, 1138

\bibitem[\protect\citeauthoryear{Haslam, Salter, Stoffel \& Wilson}{Haslam
  et~al.}{1982}]{Haslam:1982zz}
Haslam C.,  Salter C.,  Stoffel H.,    Wilson W.,  1982,
  astron.astrophys.Suppl.Ser., 47, 1

\bibitem[\protect\citeauthoryear{Huterer, Knox \& Nichol}{Huterer
  et~al.}{2001}]{Huterer:2000uj}
Huterer D.,  Knox L.,    Nichol R.~C.,  2001, astrophys.J., 555, 547

\bibitem[\protect\citeauthoryear{Hyv{\"a}rinen}{Hyv{\"a}rinen}{1999}]{DBLP:journals/tnn/Hyvarinen99}
Hyv{\"a}rinen A.,  1999, IEEE Transactions on Neural Networks, 10, 626

\bibitem[\protect\citeauthoryear{Jarvis \& Rawlings}{Jarvis \&
  Rawlings}{2004}]{Jarvis:2004gh}
Jarvis M.~J.,  Rawlings S.,  2004, New astron.Rev., 48, 1173

\bibitem[\protect\citeauthoryear{Jelic, Zaroubi, Labropoulos, Thomas, Bernardi
  et~al.,}{Jelic et~al.}{2008}]{Jelic:2008jg}
Jelic V.,  Zaroubi S.,  Labropoulos P.,  Thomas R.~M.,  Bernardi G.,    et~al.,
  2008, Mon.Not.Roy.astron.Soc., 389, 1319

\bibitem[\protect\citeauthoryear{Lewis, Challinor \& Lasenby}{Lewis
  et~al.}{2000}]{Lewis:1999bs}
Lewis A.,  Challinor A.,    Lasenby A.,  2000, astrophys.J., 538, 473

\bibitem[\protect\citeauthoryear{{Lidz}, {Furlanetto}, {Oh}, {Aguirre},
  {Chang}, {Dor{\'e}} \& {Pritchard}}{{Lidz}
  et~al.}{2011}]{2011ApJ...741...70L}
{Lidz} A.,  {Furlanetto} S.~R.,  {Oh} S.~P.,  {Aguirre} J.,  {Chang} T.-C.,
  {Dor{\'e}} O.,    {Pritchard} J.~R.,  2011, astrophys.J., 741, 70

\bibitem[\protect\citeauthoryear{{Liu} \& {Tegmark}}{{Liu} \&
  {Tegmark}}{2011}]{2011PhRvD..83j3006L}
{Liu} A.,  {Tegmark} M.,  2011, Phys.Rev., 83, 103006

\bibitem[\protect\citeauthoryear{{Liu} \& {Tegmark}}{{Liu} \&
  {Tegmark}}{2012}]{2012MNRAS.419.3491L}
{Liu} A.,  {Tegmark} M.,  2012, Mon.Not.Roy.astron.Soc., 419, 3491

\bibitem[\protect\citeauthoryear{{Liu}, {Tegmark}, {Bowman}, {Hewitt} \&
  {Zaldarriaga}}{{Liu} et~al.}{2009}]{2009MNRAS.398..401L}
{Liu} A.,  {Tegmark} M.,  {Bowman} J.,  {Hewitt} J.,    {Zaldarriaga} M.,
  2009, Mon.Not.Roy.astron.Soc., 398, 401

\bibitem[\protect\citeauthoryear{{Maino}, {Farusi}, {Baccigalupi}, {Perrotta},
  {Banday}, {Bedini}, {Burigana}, {De Zotti}, {G{\'o}rski} \&
  {Salerno}}{{Maino} et~al.}{2002}]{Maino:2001vz}
{Maino} D.,  {Farusi} A.,  {Baccigalupi} C.,  {Perrotta} F.,  {Banday} A.~J.,
  {Bedini} L.,  {Burigana} C.,  {De Zotti} G.,  {G{\'o}rski} K.~M.,
  {Salerno} E.,  2002, Mon.Not.Roy.astron.Soc., 334, 53

\bibitem[\protect\citeauthoryear{{Masui}, {Switzer}, {Banavar} N.~and{Bandura},
  {Blake}, {Calin}, {Chang}, {Chen}, {Li}, {Liao}, {Natarajan}, {Pen},
  {Peterson}, {Shaw} \& {Voytek}}{{Masui} et~al.}{2013}]{2013ApJ...763L..20M}
{Masui} K.~W.,  {Switzer} E.~R.,  {Banavar} N.~and{Bandura} K.,  {Blake} C.,
  {Calin} L.-M.,  {Chang} T.-C.,  {Chen} X.,  {Li} Y.-C.,  {Liao} Y.-W.,
  {Natarajan} A.,  {Pen} U.-L.,  {Peterson} J.~B.,  {Shaw} J.~R.,    {Voytek}
  T.~C.,  2013, astrophys.J.Lett., 763, L20

\bibitem[\protect\citeauthoryear{Mo \& White}{Mo \& White}{1996}]{Mo:1995cs}
Mo H.,  White S.~D.,  1996, Mon.Not.Roy.astron.Soc., 282, 347

\bibitem[\protect\citeauthoryear{{Moore}, {Aguirre}, {Parsons}, {Jacobs} \&
  {Pober}}{{Moore} et~al.}{2013}]{2013ApJ...769..154M}
{Moore} D.~F.,  {Aguirre} J.~E.,  {Parsons} A.~R.,  {Jacobs} D.~C.,    {Pober}
  J.~C.,  2013, astrophys.J., 769, 154

\bibitem[\protect\citeauthoryear{Morales, Bowman \& Hewitt}{Morales
  et~al.}{2006}]{Morales:2005qk}
Morales M.~F.,  Bowman J.~D.,    Hewitt J.~N.,  2006, astrophys.J., 648, 767

\bibitem[\protect\citeauthoryear{Oh \& Mack}{Oh \& Mack}{2003}]{Oh:2003jy}
Oh S.~P.,  Mack K.~J.,  2003, Mon.Not.Roy.astron.Soc., 346, 871

\bibitem[\protect\citeauthoryear{{Peebles}}{{Peebles}}{1973}]{1973ApJ...185..413P}
{Peebles} P.~J.~E.,  1973, 185, 413

\bibitem[\protect\citeauthoryear{Percival et~al.,}{Percival
  et~al.}{2001}]{Percival:2001hw}
Percival W.~J.,  et~al., 2001, Mon.Not.Roy.astron.Soc., 327, 1297

\bibitem[\protect\citeauthoryear{Peterson \& Suarez}{Peterson \&
  Suarez}{2012}]{Peterson:2012hb}
Peterson J.,  Suarez E.,  2012

\bibitem[\protect\citeauthoryear{Peterson, Aleksan, Ansari, Bandura, Bond
  et~al.,}{Peterson et~al.}{2009}]{Peterson:2009ka}
Peterson J.~B.,  Aleksan R.,  Ansari R.,  Bandura K.,  Bond D.,    et~al., 2009

\bibitem[\protect\citeauthoryear{{Petrovic} \& {Oh}}{{Petrovic} \&
  {Oh}}{2011}]{2011MNRAS.413.2103P}
{Petrovic} N.,  {Oh} S.~P.,  2011, Mon.Not.Roy.astron.Soc., 413, 2103

\bibitem[\protect\citeauthoryear{Press \& Schechter}{Press \&
  Schechter}{1974}]{Press:1973iz}
Press W.~H.,  Schechter P.,  1974, astrophys.J., 187, 425

\bibitem[\protect\citeauthoryear{{Pritchard} \& {Loeb}}{{Pritchard} \&
  {Loeb}}{2012}]{2012RPPh...75h6901P}
{Pritchard} J.~R.,  {Loeb} A.,  2012, Reports on Progress in Physics, 75,
  086901

\bibitem[\protect\citeauthoryear{{Pullen}, {Dore} \& {Bock}}{{Pullen}
  et~al.}{2013}]{2013arXiv1309.2295P}
{Pullen} A.,  {Dore} O.,    {Bock} J.,  2013, ArXiv e-prints

\bibitem[\protect\citeauthoryear{Santos, Cooray \& Knox}{Santos
  et~al.}{2005}]{Santos:2004ju}
Santos M.~G.,  Cooray A.,    Knox L.,  2005, astrophys.J., 625, 575

\bibitem[\protect\citeauthoryear{{Shaw}, {Sigurdson}, {Pen}, {Stebbins} \&
  {Sitwell}}{{Shaw} et~al.}{2013}]{Shaw2013}
{Shaw} J.~R.,  {Sigurdson} K.,  {Pen} U.-L.,  {Stebbins} A.,    {Sitwell} M.,
  2013, ArXiv e-prints

\bibitem[\protect\citeauthoryear{Sheth \& Tormen}{Sheth \&
  Tormen}{1999}]{Sheth:1999mn}
Sheth R.~K.,  Tormen G.,  1999, Mon.Not.Roy.astron.Soc., 308, 119

\bibitem[\protect\citeauthoryear{{Sullivan}, {Mobasher}, {Chan}, {Cram},
  {Ellis}, {Treyer} \& {Hopkins}}{{Sullivan}
  et~al.}{2001}]{2001ApJ...558...72S}
{Sullivan} M.,  {Mobasher} B.,  {Chan} B.,  {Cram} L.,  {Ellis} R.,  {Treyer}
  M.,    {Hopkins} A.,  2001, astrophys.J., 558, 72

\bibitem[\protect\citeauthoryear{{Switzer}, {Masui}, {Bandura}, {Calin},
  {Chang}, {Chen}, {Li}, {Liao}, {Natarajan}, {Pen}, {Peterson}, {Shaw} \&
  {Voytek}}{{Switzer} et~al.}{2013}]{2013MNRAS.434L..46S}
{Switzer} E.~R.,  {Masui} K.~W.,  {Bandura} K.,  {Calin} L.-M.,  {Chang} T.-C.,
   {Chen} X.-L.,  {Li} Y.-C.,  {Liao} Y.-W.,  {Natarajan} A.,  {Pen} U.-L.,
  {Peterson} J.~B.,  {Shaw} J.~R.,    {Voytek} T.~C.,  2013,
  Mon.Not.Roy.astron.Soc., 434, L46

\bibitem[\protect\citeauthoryear{Tegmark et~al.,}{Tegmark
  et~al.}{2004}]{Tegmark:2003uf}
Tegmark M.,  et~al., 2004, astrophys.J., 606, 702

\bibitem[\protect\citeauthoryear{Tegmark et~al.,}{Tegmark
  et~al.}{2006}]{Tegmark:2006az}
Tegmark M.,  et~al., 2006, Phys.Rev., D74, 123507

\bibitem[\protect\citeauthoryear{Tegmark, Hamilton \& Xu}{Tegmark
  et~al.}{2002}]{Tegmark:2001jh}
Tegmark M.,  Hamilton A.~J.,    Xu Y.-Z.,  2002, Mon.Not.Roy.astron.Soc., 335,
  887

\bibitem[\protect\citeauthoryear{{Testori}, {Reich}, {Bava}, {Colomb},
  {Hurrel}, {Larrarte}, {Reich} \& {Sanz}}{{Testori}
  et~al.}{2001}]{Testori:2001vp}
{Testori} J.~C.,  {Reich} P.,  {Bava} J.~A.,  {Colomb} F.~R.,  {Hurrel} E.~E.,
  {Larrarte} J.~J.,  {Reich} W.,    {Sanz} A.~J.,  2001,
  Astronomy\&Astrophysics, 368, 1123

\bibitem[\protect\citeauthoryear{{Thomas}, {Abdalla} \& {Lahav}}{{Thomas}
  et~al.}{2011}]{2011PhRvL.106x1301T}
{Thomas} S.~A.,  {Abdalla} F.~B.,    {Lahav} O.,  2011, Physical Review
  Letters, 106, 241301

\bibitem[\protect\citeauthoryear{van Haarlem, Wise, Gunst, Heald, McKean
  et~al.,}{van Haarlem et~al.}{2013}]{vanHaarlem:2013dsa}
van Haarlem M.,  Wise M.,  Gunst A.,  Heald G.,  McKean J.,    et~al., 2013,
  Astronomy\&Astrophysics, 556, A2

\bibitem[\protect\citeauthoryear{Visbal, Trac \& Loeb}{Visbal
  et~al.}{2011}]{Visbal:2011ee}
Visbal E.,  Trac H.,    Loeb A.,  2011, JCAP, 1108, 010

\bibitem[\protect\citeauthoryear{{Vujanovic}, {Staveley-Smith}, {Pen}, {Chang}
  \& {Peterson}}{{Vujanovic} et~al.}{2009}]{2009atnf.prop.2491V}
{Vujanovic} G.,  {Staveley-Smith} L.,  {Pen} U.-L.,  {Chang} T.-C.,
  {Peterson} J.,  2009, ATNF Proposal, p.~2491

\bibitem[\protect\citeauthoryear{Wang, Tegmark, Santos \& Knox}{Wang
  et~al.}{2006}]{Wang:2005zj}
Wang X.-M.,  Tegmark M.,  Santos M.,    Knox L.,  2006, astrophys.J., 650, 529

\bibitem[\protect\citeauthoryear{Willott, Rawlings, Blundell, Lacy \&
  Eales}{Willott et~al.}{2001}]{Willott:2000dh}
Willott C.~J.,  Rawlings S.,  Blundell K.~M.,  Lacy M.,    Eales S.~A.,  2001,
  Mon.Not.Roy.astron.Soc., 322, 536

\bibitem[\protect\citeauthoryear{Wilman, Miller, Jarvis, Mauch, Levrier
  et~al.,}{Wilman et~al.}{2008}]{Wilman:2008ew}
Wilman R.,  Miller L.,  Jarvis M.,  Mauch T.,  Levrier F.,    et~al., 2008,
  Mon.Not.Roy.astron.Soc., 388, 1335

\bibitem[\protect\citeauthoryear{Wyithe, Loeb \& Geil}{Wyithe
  et~al.}{2007}]{Wyithe:2007rq}
Wyithe S.,  Loeb A.,    Geil P.,  2007, Mon.Not.Roy.astron.Soc.

\bibitem[\protect\citeauthoryear{Yun, Reddy \& Condon}{Yun
  et~al.}{2001}]{Yun:2001jx}
Yun M.~S.,  Reddy N.~A.,    Condon J.~J.,  2001, astrophys.J., 554, 803

\bibitem[\protect\citeauthoryear{Zwaan et~al.,}{Zwaan
  et~al.}{2003}]{Zwaan:2003hp}
Zwaan M.~A.,  et~al., 2003, astron.J., 125, 2842

\end{thebibliography}

\appendix

\label{lastpage}

\end{document}